\documentclass[notitlepage,twoside,9pt]{article}
\usepackage[twoside,top=20truemm,bottom=20truemm,inner=20truemm,outer=20truemm]{geometry}
\usepackage{microtype}
\usepackage{color}
\usepackage[dvipsnames,svgnames,x11names]{xcolor}
\usepackage{eso-pic}

\usepackage[explicit]{titlesec}
\titleformat{\part}{\centering\LARGE\bfseries}{\thepart}{1em}{#1}[]
\titleformat{name=\part,numberless}{\centering\LARGE\bfseries}{}{1em}{#1}[]
\titleformat{\section}{\Large}{}{1em}{\titleline[l]{\rule[-4pt]{0.2cm}{1.25em}\hspace{0.3em}\thesection\hspace{1em}#1}\titlerule}[]
\titleformat{name=\section,numberless}{\Large}{}{1em}{\titleline[l]{\rule[-4pt]{0.2cm}{1.25em}\hspace{0.3em}#1}\titlerule}[]
\titleformat{\subsection}{\large}{}{1em}{\titleline[l]{\rule[-4pt]{0.2cm}{1.25em}\hspace{0.3em}\thesubsection\hspace{1em}#1}\titlerule}[]
\titleformat{name=\subsection,numberless}{\large}{}{1em}{\titleline[l]{\rule[-4pt]{0.2cm}{1.25em}\hspace{0.3em}#1}\titlerule}[]
\titleformat{\subsubsection}{\normalsize}{}{1em}{\titleline[l]{\rule[-4pt]{0.2cm}{1.25em}\hspace{0.3em}\thesubsubsection\hspace{1em}#1}\titlerule}[]
\titleformat{name=\subsubsection,numberless}{\normalsize}{}{1em}{\titleline[l]{\rule[-4pt]{0.2cm}{1.25em}\hspace{0.3em}#1}\titlerule}[]
\titleformat{\paragraph}[runin]{\normalsize\bfseries}{}{0em}{\hspace{0.5em}#1}
\usepackage[T1]{fontenc}
\usepackage{float}
\usepackage{amsmath,amssymb,amsfonts,amsthm}
\usepackage{accents}
\usepackage{mathrsfs}
\usepackage{mathtools}
\usepackage{manfnt}
\usepackage{textgreek}
\usepackage{wrapfig}

\usepackage{bm}
\usepackage{bbm}
\usepackage{dsfont}
\usepackage{yfonts}
\usepackage[bbgreekl]{mathbbol}
\usepackage{eucal}
\usepackage{halloweenmath}
\usepackage{upgreek}
\usepackage{accents}
\usepackage{blkarray}
\usepackage{multirow}
\usepackage{multicol}
\usepackage[cc]{titlepic}
\usepackage{hyperref}
\usepackage{graphicx}
\mathtoolsset{showonlyrefs=true,showmanualtags=true}
\usepackage{utfsym}
\usepackage{hwemoji}
\usepackage{hieroglf}
\usepackage{tikzbricks}
\numberwithin{equation}{section}
\usepackage{braket}
\usepackage{cancel}
\usepackage{slashed}
\usepackage{shadowtext}
\usepackage{contour}
\usepackage{niceframe}
\usepackage{fancybox}
\usepackage{fancyhdr}
\usepackage{psvectorian}
\usepackage{pgfornament}
\usepackage{tree}
\usepackage{annotate-equations}
\usepackage{scsnowman}
\usepackage{epsdice}
\usepackage{lipsum,multicol}
\usepackage[geometry,electronic,weather,misc,clock,alpine]{ifsym}
\usepackage{stmaryrd}
\usepackage{slashed}
\usepackage[inline]{enumitem}
\usepackage[all]{xy}
\usepackage{tensor}
\theoremstyle{definition}
\newtheorem{othm}{Theorem}[section]
\newenvironment{thm}
{\begin{othm}}{\hfill---\end{othm}}
\newtheorem*{thm*}{Theorem}
\theoremstyle{definition}
\newtheorem{odfn}{Definition}[section]
\newenvironment{dfn}
{\begin{odfn}}{\hfill---\end{odfn}}
\usepackage[cc]{titlepic}
\newtheorem{oprp}{Proposition}[section]
\newenvironment{prp}
{\begin{oprp}}{\hfill---\end{oprp}}
\newtheorem*{prp*}{Proposition}
\newtheorem{oexm}{Example}[section]
\newenvironment{exm}
{\begin{oexm}}{\hfill---\end{oexm}}
\newtheorem*{exm*}{Example}
\newtheorem{ocor}{Corollary}[section]
\newenvironment{cor}
{\begin{ocor}}{\hfill---\end{ocor}}
\newtheorem*{cor*}{Corollary}
\newtheorem{olem}{Lemma}[section]
\newenvironment{lem}
{\begin{olem}}{\hfill---\end{olem}}
\newtheorem*{lem*}{Lemma}
\newtheorem{orem}{Remark}[section]
\newenvironment{rem}
{\begin{orem}}{\hfill---\end{orem}}
\newtheorem*{rem*}{Remark}

\usepackage{tikz}
\usetikzlibrary{cd}
\usetikzlibrary{intersections,calc,arrows,tikzmark}
\usetikzlibrary{arrows.meta}
\usetikzlibrary{bending}
\usetikzlibrary{decorations.pathreplacing}
\usetikzlibrary{decorations.pathmorphing}
\usetikzlibrary{decorations.text}
\usetikzlibrary{decorations.markings}
\usetikzlibrary{shapes}
\usetikzlibrary{shadings}
\usetikzlibrary{patterns.meta}
\usetikzlibrary{backgrounds,fit}
\usepackage{tikzsymbols}
\usepackage[breakable]{tcolorbox}
\tcbuselibrary{breakable,theorems,skins,xparse,vignette,hooks}
\tcbset{ghostbox/.style={enhanced,frame hidden,interior hidden,size=minimal,remember as=#1}}
\usepackage{postit}
\definecolor{mytext}{rgb}{0.8,0.9,0.95}
\definecolor{mypage}{rgb}{0,0.2,0.3}
\definecolor{myapage}{rgb}{0,0.3,0.4}
\definecolor{anothermytext}{rgb}{0.7,0.2,0.5}
\definecolor{bb}{rgb}{0.25,0.35,0.4}
\definecolor{bbwhite}{rgb}{0.87,0.99,0.99}
\definecolor{colitem}{rgb}{0,0.7,0.9}
\definecolor{sum1}{rgb}{0,0.5,0.5}
\definecolor{sum2}{rgb}{0.5,0,0.5}
\definecolor{sum3}{rgb}{0.5,0.5,0}
\definecolor{sum4}{rgb}{0,0.75,0.5}
\definecolor{sum5}{rgb}{0.5,0,0.75}
\definecolor{sum6}{rgb}{0.75,0.5,0}
\definecolor{sum7}{rgb}{0,0.5,0.75}
\definecolor{sum8}{rgb}{0.75,0,0.5}
\definecolor{sum9}{rgb}{0.5,0.75,0}
\definecolor{chalkboard}{rgb}{0,0.25,0}
\definecolor{pagecolor1}{rgb}{0,0.15,0.25}
\definecolor{pagecolor2}{rgb}{0,0.2,0.3}
\definecolor{pagecolor3}{rgb}{0,0.3,0.4}
\definecolor{columntext}{rgb}{0,0.2,0.2}
\hypersetup{colorlinks=true,linkcolor=RoyalBlue1,citecolor=Emerald}

\tcolorboxenvironment{myitemize}{blanker,before skip=15pt,after skip=15pt,borderline west={1mm}{0pt}{PaleTurquoise3!70!PaleGreen1},colupper=mytext}
\DeclareMathAlphabet{\omathbb}{U}{msb}{m}{n}
\DeclareMathAlphabet{\matheur}{U}{eur}{m}{n}

\DeclareMathOperator{\tr}{tr}

\DeclareMathOperator{\Ad}{Ad}

\let\O\relax
\newcommand{\O}{\mathrm{O}}
\newcommand{\SO}{\mathrm{SO}}
\newcommand{\ISO}{\mathrm{ISO}}

\newcommand{\so}{\mathfrak{so}}
\newcommand{\iso}{\mathfrak{iso}}

\DeclarePairedDelimiter{\rbra}{\lparen}{\rparen}
\DeclarePairedDelimiter{\cbra}{\lbrace}{\rbrace}
\DeclarePairedDelimiter{\sbra}{\lbrack}{\rbrack}

\DeclarePairedDelimiter{\abra}{\langle}{\rangle}

\DeclarePairedDelimiterX{\intset}[2]{\lbrace}{\rbrace}{#1\,\delimsize\vert\,#2}
\DeclarePairedDelimiterX{\rpair}[2]{\lparen}{\rparen}{#1,#2}
\DeclarePairedDelimiterX{\cpair}[2]{\lbrace}{\rbrace}{#1,#2}
\DeclarePairedDelimiterX{\spair}[2]{\lbrack}{\rbrack}{#1,#2}
\DeclarePairedDelimiterX{\wedgespair}[2]{\lbrack}{\rbrack}{#1\mathpunct{\overset{\wedge}{,}}#2}
\DeclarePairedDelimiterX{\starspair}[2]{\lbrack}{\rbrack}{#1\mathpunct{\overset{\star}{,}}#2}
\DeclarePairedDelimiterX{\starwedgespair}[2]{\lbrack}{\rbrack}{#1\mathpunct{\overset{\overset{\star}{\wedge}}{,}}#2}
\DeclarePairedDelimiterX{\cjac}[3]{\lbrace}{\rbrace}{#1,\lbrace#2,#3\rbrace}
\DeclarePairedDelimiterX{\sjac}[3]{\lbrack}{\rbrack}{#1,\lbrack#2,#3\rbrack}
\DeclarePairedDelimiterX{\rset}[2]{\lparen}{\rparen}{#1,\dots,#2}
\DeclarePairedDelimiterX{\cset}[2]{\lbrace}{\rbrace}{#1,\dots,#2}
\DeclarePairedDelimiterX{\sset}[2]{\lbrack}{\rbrack}{#1,\dots,#2}

\usepackage{tikzsymbols}
\usepackage{wasysym}

\newcommand{\quot}[2]{{^{\displaystyle{#1}}}\hspace{-0.5mm}\Big/\hspace{-0.5mm}{_{\displaystyle{#2}}}}

\newtcolorbox{column}[1][]{enhanced,breakable,colback=mytext,colframe=mytext,coltitle=columntext,coltext=columntext,boxrule=1pt,fonttitle=\bfseries,title=Column : {#1},frame style={decorate,decoration={random steps,segment length=3mm,amplitude=0.7mm}}}

\makeatletter

\@addtoreset{figure}{section}

\@addtoreset{table}{section}
\makeatother
\makeatletter
\def\@oddhead{\thesubsection\hfil\thepage}
\makeatletter
\def\@fnsymbol#1{\ensuremath{\ifcase#1\or\dagger\else\@ctrerr\fi}}
\makeatother
\color[rgb]{0,0,0}
\pagecolor[rgb]{0.999,0.999,0.999}
\title{Higher-form Symmetries of Linearized Gravity\\in Maximally Symmetric Spaces}
\author{Natsuya Kido\thanks{Department of Physics, Ibaraki University, Mito 310-8512, Japan}}
\date{\today}
\begin{document}
\maketitle
\begin{abstract}
We formulate the gravitational field on maximally symmetric spacetimes as a gauge theory, and investigate higher-form symmetries of the linearized gravitational field via a series expansion in Newton's gravitational constant. To treat gravity as a gauge theory, the MacDowell-Mansouri formulation based on Cartan geometry is utilized. We show the existence of a $1$-form symmetry in linearized gravity on Minkowski spacetime. Furthermore, for other maximally symmetric spacetimes (de Sitter and anti-de Sitter spacetimes) with a small cosmological constant, we demonstrate the existence of an approximate $1$-form symmetry in the corresponding linearized gravity, detailing its explicit dependence on the cosmological constant.
\end{abstract}
\vspace{-0.7cm}
\tableofcontents

\newpage
\section{Introduction}
In recent years, the concept of generalized symmetries has attracted significant attention in theoretical physics. Initiated by Gaiotto, Kapustin, Seiberg, and Willett~\cite{Gaiotto-Kapustin-Seiberg-Willett}, this notion has substantially broadened the conventional framework of symmetry. On the one hand, the charged operators under transformation are extended from local operators to extendes ones; on the other hand, the algebraic structure governing the product of symmetry-generating operators is generalized beyond groups. Below, we briefly introduce the former extension, which is the main focus of this paper.

Such generalized symmetries constitute an important research topic in modern quantum field theory. However, this conceptual framework is by imported into other areas of physics. In this regard, the field-theoretic techniques developed in the present work may serve as a useful guide for future theoretical developments in those domains.

The central subject of this paper is the analysis of gravitational theories formulated in terms of the vielbein and spin connection, through the lens of generalized symmetries.

\vspace{\baselineskip}

In Maxwell electromagnetism, the conservation of electric charge means, in short, that the electric current satisfies the continuity equation (at the classical level). The magnitude of the electric current is measured along curves in spacetime, thus, the electric current can be described as a $1$-form on spacetime. Alternatively, it can be formulated equivalently using its Hodge dual (which becomes a $3$-form in four-dimensional spacetime). When described using differential forms, the continuity equation corresponds to the condition that either the codifferential or the exterior derivative of the electric current vanishes (on-shell). Which derivative vanishes depends on whether the current is treated as a $1$-form or as its Hodge dual, and since both representations are physically identical, choosing between them is a matter of preference. Here, we adopt the former perspective, in which the current is defined as a $1$-form whose vanishing codifferential constitutes the continuity equation.

Incidentally, within the framework of Maxwell electromagnetism, there exist differential forms other than the electric current whose codifferential or exterior derivative vanishes, In the absence of source electric charges and currents, the electromagnetic field in Minkowski or Euclidean spacetime---which is globally defined $2$-form $F$ on these spacetime---satisfies the Maxwell equation and the Bianchi identity:
\begin{align}
d^{\dagger}F=0,\qquad dF=0,
\end{align}
where $d$ denotes the exterior derivative and $d^{\dagger}$ denotes the codifferential. Among these, the Maxwell equation represents the equation of motion for the electromagnetic field and therefore holds only on-shell. In contrast, the Bianchi identity is a geometric identity that holds universally when the electromagnetic field is described in this manner. Thus, although the conditions under which they hold differ, this theory contains $2$-form $F$ that is coclosed or closed---that is, a form whose codifferential or exterior derivative vanishes under specific conditions.

If the continuity equation signifies the (on-shell) coclosedness of the current, then conversely, by interpreting any coclosed differential form---not limited to $1$-forms---as a conserved current satisfying a continuity equation, the concept of symmetry underlying such conservation laws is naturally extended. Although this viewpoint is based on classical theory, an analogous argument can be made at the quantum level by replacing the continuity equation with the Ward-Takahashi identity. In any case, this generalized concept of symmetry, which regards higher-degree differential forms beyond $1$-forms as conserved currents, is called ``higher-form symmetry'' and constitutes one of the important reserch fields in modern quantum field theory.

In the present example, the electromagnetic field $F$ serves as a $2$-form conserved current, and the corresponding conservation of charges reflects the conservation of electric and magnetic flux in regions free of electric charges and currents. A symmetry equipped with a $2$-form conserved current is referred to as a $1$-form symmetry (whereas conventional symmetries are understood as $0$-form symmetries). Furthermore, in Minkowski (or Euclidean) spacetime with trivial de Rham cohomology, the closed form $F$ implies the existence of a globally defined photon $A$ satisfying $F=dA$. Consequently, the current becomes (globally) an exact form. This field $A$ can be interpreted as the Nambu-Goldstone boson associated with the higher-form symmetry (in this case, the $1$-form symmetry), which means that within the context of extended symmetries, the photon itself is naturally viewed as a Nambu-Goldstone boson.

\vspace{\baselineskip}

Turning our attention from electromagnetism to gravitational theories, one notices the existence of a structure analogous to a classical gauge field. Naturally, numerous differences exist between the two, as well as fundamental distinctions from a quantum-theoretical perspective, preventing a complerely parallel treatment. Nevertheless, they share common properties, and focusing on these shared aspects to exchange insights and find mutual applications is a worthwhile endeavor. In this study, we focus on gravity in the regime where its influence is sufficiently weak compared to other fundamental interactions---namely, linearized gravity---and discuss it by analogy with electromagnetic theory. Regarding linearized gravity obtained by linearizing classical general relativity, whether its quantization coincides with the linearization of a true quantum theory of gravity---including the very question of whether a consistent framework of linearized quantum gravity exists---remains a challenging problem beyond our present scope. Leaving aside this issue regarding the commutativity of linearization and quantization for gravitational fields, we consider a quantum field theory constructed by quantizing classical linearized gravity. Although there is no guarantee that this corresponds to the linearization of quantum gravity, we proceed under the working hypothesis that this quantized field provides meaningful insights into quantum gravity.

In this study, we discuss higher-form symmetries in linearized gravity, with a particular focus on the shift symmetry of the gauge fields observed in Maxwell electromagnetism. A related study on this topic was conducted by Hull, Hutt, and Lindstr\"{o}m~\cite{Hull-Hutt-Lindstr\"om}. While their analysis was carried out using the metric tensor, we adopt a different perspective in this work by formulating the discussion in terms of the vielbein and spin connection. Furthermore, by employing the MacDowell-Mansouri formulation of gravity~\cite{MacDowell-Mansouri}, we extend our analysis of higher-form symmetries to naturally incorporate de Sitter and anti-de Sitter spacetime.

The remainder of this paper is organized as follows. First, we examine the higher-form symmetries present in linearized gravity on Minkowski spacetime (Section~\ref{sec:1FSoLGoMST}). Subsequently, we extend our discussion of higher-form symmetries to incorporate de Sitter and anti-de Sitter spacetimes (Section~\ref{sec:1FSoLGfSUC}) by utilizing the MacDowell-Mansouri formulation of gravity.

The MacDowell-Mansouri formulation of gravity rests on the geometric foundation of Cartan geometry~\cite{Wise1,Wise2,Wise3}. Therefore, prior to discussing higher-form symmetries within this theory, we provide a concise review of the essential aspects of Cartan geometry (Section~\ref{sec:CG}).

\section{$1$-form symmetry of linearized gravity with zero cosmological constant}
\label{sec:1FSoLGoMST}
In this section, we consider the gravitational field in Minkowski spacetime, focusing particularly on the higher-form symmetries of linearized gravity, where its effects are relatively small. The case with a non-zero cosmological constant will be discussed in Section \ref{sec:1FSoLGfSUC}.

First, let us clarify some terminology. Given a spacetime manifold $M$, its tangent bundle $p\colon TM\rightarrow M$ is naturally defined. A bundle map $e\colon TM\rightarrow\CMcal{T}$ to vector bundle $\pi\colon\CMcal{T}\rightarrow M$---which has the same dimension as $TM$ and whose fibers are inner product spaces---is called a vielbein. Being a bundle map means that $e$ makes the following diagram commute:
\begin{center}
\begin{tikzcd}
TM\arrow[dr,"p"']\arrow[rr,"e"]&&\CMcal{T}\arrow[dl,"\pi"]\\
&M&
\end{tikzcd}
\end{center}
In \cite{Wise2}, $\CMcal{T}$ is referred to as a fake tangent bundle. Leaving its concrete picture for later, let us briefly examine what further properties can be deduced at this stage.

Each fiber of  $\CMcal{T}$ is an inner product space. We denote by $\eta_{x}$ the metric equipped on the fiber $\pi^{-1}\rbra{x}$ at an arbitrary point $x\in M$, and let $\eta$ be the smooth collection of these metrics across all points of $M$. Then, $\eta$ is a symmetric bilinear map that assigns a smooth function to any pair of smooth sections of $\CMcal{T}$. Pulling this back by $e$, we obtain a pseudo-Riemannian metric $g=e^{\ast}\eta$ on $M$. Typically, $\eta$ is the Minkowski metric, and its pullback $g$ becomes a Lorentzian metric. Of course, depending on the context, one may choose $\eta$ to be a Euclidean metric, in which case $g$ becomes a Riemannian metric. The choice of admissible metrics $\eta$ depends on the structure of the vector bundle chosen for $\CMcal{T}$. Next, let us make $\CMcal{T}$ concrete to explore this futher.

Let the dimension of $M$ be $d+1$. As the fake tangent bundle $\CMcal{T}$, let us consider the associated vector bundle $\mathscr{F}\times^{\SO\rbra{d,1}}_{\Ad}\omathbb{R}^{d,1}$ of the $\SO\rbra{d,1}$ frame bundle $\mathscr{F}\rightarrow M$. The typical fiber $\omathbb{R}^{d,1}$ is a $\rbra{d+1}$-dimensional Minkowski space equipped with the Minkowski metric $\eta$. Combined with the above discussion, a general spacetime metric is obtained from this $\eta$ via the vielbein. Furthermore, the pullback $\omega$ of Ehresmann connection on $\mathscr{F}$ to $M$ by a section is called a spin connection. This defines a connection on $\mathscr{F}\times^{\SO\rbra{d,1}}_{\Ad}\omathbb{R}^{d,1}$, and pulling it back futher by the vielbein yields the Levi-Civita connection on $M$.

\vspace{\baselineskip}

Under this setup, we shall investigate the action functional that governs the dynamics of the vielbein $e$ and the spin connection $\omega$, its linearization, and the higher-form symmetries that emerge therein.

\subsection{Palatini action functional}
In the remainder of this section, we assume the spacetime dimension to be $3+1$.

Let $\omega$ and $e$ be the spin connection and the vielbein on a $\rbra{3+1}$-dimensional Minkowski spacetime $M$, respectively. Here, $\omega$ is an $\so\rbra{3,1}$-valued $1$-form on $M$, and $e$ is an $\omathbb{R}^{3,1}$-valued $1$-form on $M$. The Palatini action functional governing these fields is given by
\begin{align}
\label{eq:PAF}
S\sbra{\omega,e}=\frac{1}{2G}\int_{M}{\star_{\text{int}}}\rbra{e\wedge e\wedge R}.
\end{align}
A few remarks are in order regarding the notation. First, $R\coloneqq d\omega+\frac{1}{2}\wedgespair{\omega}{\omega}$ is the Riemann curvature. The symbol $\wedge$ in the above equation combines the exterior product of differential forms on $M$ with the exterior product in $\so\rbra{3,1}\cong\Lambda^{2}\omathbb{R}^{3,1}$, and ${\star_{\text{int}}}$ denotes the Hodge star operator on $\Lambda^{\bullet}\omathbb{R}^{3,1}$. Furthermore, while the exterior derivative acting on differential forms on $M$ appears frequently throughout this paper, the exterior derivative on $\Lambda^{\bullet}\omathbb{R}^{3,1}$ does not appear. Therefore, $d$ exclusively denotes the exterior derivative acting on differential forms on $M$. Here, $G$ denotes Newton's gravitational constant.

Varying the Palatini action functional with respect to the vielbein $e$ yields the equation of motion:
\begin{align}
e\wedge R=0.
\end{align}
When the vielbein $e$ is a bundle isomorphism---that is, non-degenerate---this equation implies $R=0$ (in general, $R=0$ is a sufficient condition for $e\wedge R=0$, but not necessarily a necessary one). This indicates that, in the absence of energy-momentum of matter (including gauge fields other than gravity), Minkowski spacetime is flat in the sense of Riemannian geometry. On the other hand, varying the Palatini action functional with respect to the spin connection $\omega$ yields the equation of motion:
\begin{align}
d\rbra{e\wedge e}+\wedgespair{\omega}{e\wedge e}=0.
\end{align}
If $e$ is a bundle isomorphism, this equation implies $de+\wedgespair{\omega}{e}=0$ (where the left-hand side is the covariant exterior derivative of $e$ with respect to $\omega$). This states that the torsion in the sense of Riemannian geometry vanishes.

\paragraph{Vielbein shift}
Abrupt as it may seem, let us consider a shift $e+\alpha$ of the vielbein $e$ by an $\omathbb{R}^{3,1}$-valued closed $1$-form $\alpha$. Under this transformation, the equation of motion $R=0$ remains invariant. However, the underlying equation $e\wedge R=0$ is, ingeneral, not invariant (covariant), nor is the action functional invariant unless evaluated on-shell ($e\wedge R=0$). Futhermore, the other equations of motion, $de+\wedgespair{\omega}{e}=0$ and $d\rbra{e\wedge e}+\wedgespair{\omega}{e\wedge e}=0$, are also not covariant due to the terms coupling with $\omega$\footnote{If one imposes $d_{\omega}\alpha=0$ instead of $d\alpha=0$ (whew $d_{\omega}$ denotes the exterior covariant derivative with respect to $\omega$), both $R=0$ and $de+\wedgespair{\omega}{e}=0$ remain covariant. However, this condition contradicts $d_{\omega}^{2}e=R\wedge e$, which holds geometrically regardless of whether the fields are on-shell or off-shell (although on-shell, no contradiction arises since $R=0$).}. Thus, while this transformation maps solutions of the Einstein equations to other solutions, it destroys the underlying physical consistency.

Consequently, such shift symmetries for the vielbein do not exist in the full nonlinear theory (and similarly for the spin connection). In what follows, by linearizing the gravitational fields--namely the vielbein and the spin connection--we shall develop the present discussion for the regime of relatively weak gravity.

\subsection{Linearized gravity}
This completes the basic introduction, Here, we expand the spin connection and the vielbein in a power series in the square root of Newton's gravitational constant $\sqrt{G}$, and restrict our discussion up to the linear order. In this sense, we refer to the first-order terms in $\sqrt{G}$ of the spin connection and the vielbein as linearized gravity.

\vspace{\baselineskip}

We expand the spin connection $\omega$ and the vilbein $e$ in powers of $\sqrt{G}$ as
\begin{align}
\omega=\underline{\omega}+\sqrt{G}\varpi+\CMcal{O}\rbra*{\sqrt{G}^{2}},\qquad e=\underline{e}+\sqrt{G}\varepsilon+\CMcal{O}\rbra*{\sqrt{G}^{2}}.
\end{align}
Among these terms, the zeroth-order quantities correspond to the case where no eneregy-momentum is present. In particular, for Minkowski spacetime, the zeroth-order spin connection can be set to zero, and the zeroth-order vielbein can be chosen as a closed form. Using $\underline{\omega}=0$, the Riemann curvature defined by the spin connection can then be expanded in powers of $\sqrt{G}$ as
\begin{align}
R=d\omega+\frac{1}{2}\wedgespair{\omega}{\omega}=\sqrt{G}d\varpi+\sqrt{G}^{2}\rbra*{d\varpi^{\prime}+\frac{1}{2}\wedgespair{\varpi}{\varpi}}+\CMcal{O}\rbra*{\sqrt{G}^{3}},
\end{align}
where $\varpi^{\prime}$ denotes the second-order term of $\omega$ in $\sqrt{G}$.

Using the above expansions, expanding the Palatini action functional \eqref{eq:PAF} yields, at leading order,
\begin{align}
\label{eq:PAFfLG}
S\sbra{\varpi,\varepsilon}=\int_{M}{\star_{\text{int}}}\rbra*{\underline{e}\wedge\varepsilon\wedge d\varpi+\frac{1}{4}\underline{e}\wedge\underline{e}\wedge\wedgespair{\varpi}{\varpi}},
\end{align}
where the term originating from $d\varpi^{\prime}$ has been dropped via integration by parts. Since $\underline{e}$ is a background (given) quantity, this action functional governs the dynamics of $\varpi$ and $\varepsilon$.

To derive the equations of motion that govern the dynamics of $\varpi$ and $\varepsilon$, let us vary the above action functional with respect to $\varepsilon$ and $\varpi$. First, varying with respect to $\varepsilon$ yields
\begin{align}
\underline{e}\wedge d\varpi=0.
\end{align}
Assuming that $\underline{e}$ is non-degenerate (or a bundle isomorphism), this equation is equivalent to $d\varpi=0$. Next, varying with respect to $\varpi$ yields
\begin{align}
d\rbra{\underline{e}\wedge\varepsilon}+\wedgespair{\varpi}{\underline{e}\wedge\underline{e}}=0.
\end{align}
Provided that $\underline{e}$ is non-degenerate or a bundle isomorphism, this is likewise equivalent to $d\varepsilon+\wedgespair{\varpi}{\underline{e}}=0$. In the following, we assume that $\underline{e}$ is non-degenerate.

\subsection{$1$-form symmetry of linearized gravity}
\label{subsec:1FSoLGoMST}
In this subsection, we discuss the higher-form symmetries---in paticular, the $1$-form symmetry---of linearized gravity governed by the action functional \eqref{eq:PAFfLG} prepared above. This allows us to understand the structure of the higher-form symmetries present in extremely weak gravitational interactions on Minkowski spacetime, as well as why such symmetries are absent in relatively stronger gravitational interactions.

\subsubsection{Shift symmetry}
Under the transformation $\varepsilon\mapsto\varepsilon+\alpha$ of the vielbein perturbation by an $\omathbb{R}^{3,1}$-valued closed $1$-form $\alpha$ (while leaving $\varpi$ unchanged), the linearized Palatini action functional \eqref{eq:PAFfLG} transforms as
\begin{align}
S\sbra{\varpi,\varepsilon+\alpha}=S\sbra{\varpi,\varepsilon}+\int_{M}{\star_{\text{int}}}\rbra{\underline{e}\wedge\alpha\wedge d\varpi}.
\end{align}
By performing integration by parts and using the fact that $\underline{e}$ and $\alpha$ are closed, the second term on the right-hand side of this equation vanishes (assuming the boundary terms vanish)\footnote{Note that the exterior derivative $d$ commutes with the Hodge star operator ${\star_{\text{int}}}$.}. Alternatively, on-shell, it vanishes even without integration by parts due to $d\varpi=0$ (or, more generally, $\underline{e}\wedge d\varpi=0$). Therefore, the action functional is invariant under this transformation. That is, linearized gravity described by the Palatini action functional \eqref{eq:PAFfLG} possesses a $1$-form shift symmetry with respect to the vielbein pertubation.

\subsubsection{Noether current associated with the symmetry}
Let us consider the Noether current (conserved current) associated with the aforementioned $1$-form symmetry of the vielbein perturbation. When the Hodge star operator ${\star_{\text{int}}}$ on $\Lambda^{\bullet}\omathbb{R}^{3,1}$ acts on the basis of $\Lambda^{4}\omathbb{R}^{3,1}$, it yields the completely antisymmetric tensor $\epsilon_{\mu\nu\rho\sigma}$ ($\mu,\nu,\rho,\sigma=0,1,2,3$). Bearing this in mind, under the transformation $\varepsilon\mapsto\varepsilon+\alpha$ of the vielbein perturbation by an $\omathbb{R}^{3,1}$-valued closed $1$-form $\alpha$, the variation of the action functional can be rewritten as
\begin{align}
S\sbra{\varpi,\varepsilon+\alpha}-S\sbra{\varpi,\varepsilon}=\int_{M}\alpha^{\mu}\wedge d\rbra*{-\frac{1}{2}\epsilon_{\mu\nu\rho\sigma}\underline{e}^{\nu}\wedge\varpi^{\rho\sigma}}.
\end{align}
Here, the expansion coefficients of each quantity with respect to the basis of $\Lambda^{\bullet}\omathbb{R}^{3,1}$ are denoted by $\alpha^{\mu}$, $\underline{e}^{\nu}$, and $\frac{1}{2}\varpi^{\rho\sigma}$, respectively, and we have inserted $\underline{e}^{\nu}$ inside the exterior derivative using the fact that $\underline{e}^{\nu}$ is closed. This yields the $2$-form Noether current associated with this transformation:
\begin{align}
J_{\mu}={\star_{\eta}}\rbra*{\frac{1}{2}\epsilon_{\mu\nu\rho\sigma}\underline{e}^{\nu}\wedge\varpi^{\rho\sigma}}.
\end{align}
Note that ${\star_{\eta}}$ denotes the Hodge star operator defined by the Minkowski metric $\eta$ on $M$, which acts on differential forms. Consequently, $J_{\mu}$ satisfies $d{\star_{\eta}}J_{\mu}=0$ on-shell ($d\varpi=0$). Here, if we define $\tensor[^{{\star_{\text{int}}}}]{\varpi}{_{\mu\nu}}\coloneqq\frac{1}{2}\epsilon_{\mu\nu\rho\sigma}\varpi^{\rho\sigma}$ as (twice) the expansion coefficient of ${\star_{\text{int}}}\varpi$ with respect to the basis of ${\star_{\text{int}}}\varpi$ in $\Lambda^{2}\omathbb{R}^{3,1}\cong\so\rbra{3,1}$, $J_{\mu}$ can also be expressed as
\begin{align}
J_{\mu}={\star_{\eta}}\rbra*{\underline{e}^{\nu}\wedge\tensor[^{{\star_{\text{int}}}}]{\varpi}{_{\mu\nu}}}.
\end{align}
Alternatively, considering an $\omathbb{R}^{3,1}$-valued $2$-form $J$ whose expansion coefficients with respect to the basis of $\omathbb{R}^{3,1}$ are given by $J^{\mu}$ (where indices are raised and lowered using the Minkowski metric $\eta$), we can write $J={\star_{\eta}}\wedgespair{\underline{e}}{{\star_{\text{int}}}\varpi}$. This quantity $\wedgespair{\underline{e}}{{\star_{\text{int}}}\varpi}$ is deceptively similar to, yet distinct from, $\wedgespair{\underline{e}}{\varpi}$. If it were $\wedgespair{\underline{e}}{\varpi}$, it could be rewritten as $-d\varepsilon$ under the vanishing torsion condition, but that is not the case here. Nevertheless, on-shell ($d\varepsilon+\wedge{\varpi}{\underline{e}}=0$), $\wedgespair{\underline{e}}{{\star_{\text{int}}}\varpi}$ is a closed form and thus, by Poincar\'{e}'s lemma, is also an exact form on $M$. Therefore, if we define an $\omathbb{R}^{3,1}$-valued $1$-form $\widetilde{\varepsilon}$ that satisfies $d\tilde{\varepsilon}=\wedgespair{\underline{e}}{{\star_{\text{int}}}\varpi}$ on-shell, we can write $J={\star_{\eta}}d\tilde{\varepsilon}$. The difference between $\tilde{\varepsilon}$ and $\varepsilon$ stems from the difference between ${\star_{\text{int}}}\varpi$ and $\varpi$, whose components are related by $\tensor[^{{\star_{\text{int}}}}]{\varpi}{^{\mu\nu}}=\frac{1}{2}\tensor{\epsilon}{^{\mu\nu}_{\rho\sigma}}\varpi^{\rho\sigma}$. Consequently, in terms of physical information content, $\tilde{\varepsilon}$ and $\varepsilon$ can be regarded as equivalent.

\vspace{\baselineskip}

The discussion so far rested on the assumption that the state of spacetime in the absence of matter energy-momentum is Minkowski spacetime. However, even when matter energy-momentum is absent, a non-zero cosmological constant causes spacetime to deviate from Minkowski spacetime. Therefore, in what follows, we wish to extend the same line of argument to the case with a cosmological constant. For this purpose, we require a geometric framework that describes spacetime in a unified manner, including cases with a cosmological constant. Here, rather than relying on the commonly used Riemannian (or Lorentzian) geometry, we adopt Cartan geometry---a framework that encompasses it---to carry out our discussion. In the next section, we shall provide a brief, self-contained overview of Cartan geometey, keeping it to the necessary minimum.

\section{Preparation: Cartan geometry}
\label{sec:CG}
In the next section, we intend to generalize the discussion presented in the previous section. Specifically, we focus on spacetimes possessing a non-zero cosmological constant. Such spacetimes are referred to as de Sitter or anti-de Sitter spacetimes, and it is advantageous to prepare an appropriate geometric language to describe them. Although pseudo-Riemannian geometry, or Lorentzian geometry, is typically employed to describe spacetime, a more natural framework exists when a cosmological constant is present. This is Cartan geometry, and in this section, we review its fundamentals in preparation for the subsequent section.

\vspace{\baselineskip}

Although Cartan geometry is an extension of Euclidean geometry, it is instructive to begin with its precursor to elucidate the nature of this extension. In Euclidean geometry, when considering the congruence of planar figures, for example, a figure subjected to rotations and translations is identified with the original figure. That is, the central objects of study were those invariant under the action of the group formed by these transformations on the Euclidean plane. Generalizing this concept, the geometry defined by focusing on group actions on manifolds is known as Klein geometry, in which one investigates objects that remain invariant under such actions.

Meanwhile, in particle physics and gravitational theory, the Lorentz invariance of a theory constitutes a fundamental requirement. Consequently, analyzing objects that are invariant under the action of the Lorentz group is an essential element in constructing these theories. For this reason, Klein geometry serves as a crucial language in particle physics and gravitation; however, situations can arise in these theories where Klein geometry proves insufficient. A concrete example is when spacetime is warped by the presence of matter, in which case it becomes necessary to endow Klein geometry with appropriate deformations. The geometry that extends Klein geometry to accommodate such situations is called Cartan geometry.

Thus far, we have introduced Klein geometry by focusing on group actions, and subsequently arrived at Cartan geometry as an extension that accommodates deformations. However, one can also reach Cartan geometry via different route. Another obvious extension of Euclidean geometry is to allow for curvature or deformations, leading to what is known as Riemannian geometry. By extending the group action within Riemannian geometry in a manner analogous to the previous case, one again arrives at Cartan geometry. This network of extensions is summarized in the following commutative diagram:

\begin{center}
\begin{tikzcd}[row sep=2cm,column sep=3cm]
\text{Euclidean geometry}\arrow[r,"\text{generalizing curvature}"]\arrow[d,"\text{generalizing group actions}"']&\text{Riemannian geometry}\arrow[d,"\text{generalizing group actions}"]\\
\text{Klein geometry}\arrow[r,"\text{generalizing curvature}"']&\text{Cartan geometry}
\end{tikzcd}
\end{center}

In the following, we first introduce Klein geometry, based upon which we subsequently explain Cartan geometry. With applications to gravitational theory in mind, we restrict ourselves here to a minimal exposition necessary for that purpose. For a broader perspective, the reader is referred to~\cite{Sharpe},~\cite{MacKay}, or~\cite{K\"ohler}. For a physics-oriented exposition,~\cite{Wise2} is also helpful.

\subsection{Klein geometry}
Just as Riemannian geometry is a geometry that locally resembles (flat) Euclidean geometry, Cartan geometry is also locally governed by a certain type of ``flat'' geometry. In this subsection, we first explain this underlying geometry. Physically, this corresponds to considering a universe in which curvature arises solely from the cosmological constant, with no distortion induced by the energy-momentum of matter.

Let us first define Klein geometry. As mentioned above, it is characterized by a manifold and a group action on it.
\begin{dfn}[Klein geometry]
Let $X$ be a smooth manifold on which a Lie group $G$ acts smoothly and transitively on the left (letting the action map be $\triangleright\colon G\curvearrowright X$)\footnote{An action of a group $G$ on $X$ is said to be transitive if, for any two elements $x,y\in X$, there exists some $g\in G$ such that $y=g\triangleright x$.}. In this context, the pair $\rbra{X,G}$ is called a \textbf{Klein geometry} or a homogeneous space.
\end{dfn}

\begin{rem}
In the definition of a Klein geometry $\rbra{X,G}$, additional conditions are sometimes imposed. For details on such definitions, the reader is referred to~\cite{MacKay}; briefly stated, these conditions require that the action of any open subset of $G$ on any point of $X$ yields an open subset of $X$ and that there exists a point in $X$ whose $G$-orbit contains points in all connected components of $X$. A Klein geometry satisfying these conditions possesses a diffeomorphism described below, which provides a highly convenient perspective on Klein geometries. Therefore, in what follows, we assume that a Klein geometry is accompanied by these conditions. Note that $X$ is sometimes assumed to be connected (the definition in~\cite{Sharpe} essentially corresponds to this case), under which the latter condition need not be separately imposed. All specific Klein geometries treated below have a connected $X$.
\end{rem}

In a Klein geometry $\rbra{X,G}$, fix an arbitrary point $x_{0}\in X$, and let the isotropy subgroup (stabilizer) with respect to that point be $H=\intset{g\in G}{g\triangleright x_{0}=x_{0}}$\footnote{By the transitivity of the action, even if one takes an isotropy subgroup for an element of $X$ other than $x_{0}$, it is isomorphic to the isotropy subgroup $H$ associated with $x_{0}$. Therefore, the point $x_{0}$ carries no absolute significance, and we suppress explicit dependence on $x_{0}$ in the notation $H$.}. Then, the (left) coset space
\begin{align}
\quot{G}{H}\coloneqq\intset{gH}{g\in G}
\end{align}
admits the structure of a smooth manifold, yielding a diffeomorphism
\begin{align}
\begin{array}{c@{\,\,}c@{\,\,}c@{\,\,}}
\quot{G}{H}&\longrightarrow&X\\
\rotatebox{90}{$\in$}&&\rotatebox{90}{$\in$}\\
gH&\longmapsto&g\triangleright x_{0}
\end{array}.
\end{align}
This result implies that one may equivalently employ $\quot{G}{H}$ to investigate the geometry of the space $X$. Note that $G$ is the group acting on $X$ (which essentially captures the symmetries of $X$), while $H$ is a specific subgroup thereof (corresponding to the action that fixes a particular point in $X$). In other words, Klein geometry can be characterized as a geometry that highlights the symmetries of the space.

\begin{exm}[Minkowski space]
We equip the $\rbra{d+1}$-dimansional vector space $\omathbb{R}^{d+1}$ with a metric as follows. With respect to the basis $\cbra{e_{a}}_{a=0}^{d}$ of $\omathbb{R}^{d+1}$, the metric $\eta\colon\omathbb{R}^{d+1}\times\omathbb{R}^{d+1}\rightarrow\omathbb{R}$ is defined by
\begin{align}
\eta\rbra{e_{0},e_{0}}=-1,\qquad\eta\rbra{e_{i},e_{i}}=+1\text{($i=1,\dots,d$)},\qquad\text{and zero otherwise}.
\end{align}
We denote the inner product space endowed with this metric by $\omathbb{R}^{d,1}$. This is Minkowski space equipped with the Minkowski metric. For $\omathbb{R}^{d,1}$, the isometry group $\ISO\rbra{d,1}$ is the group of all transformations that preserve the inner product of vector differences defined by the metric, acting smoothly and transitively on $\omathbb{R}^{d,1}$. Its closed subgroup, the Lorentz group $\SO\rbra{d,1}$, constitutes the isotropy subgroup woth respect to the origin of $\omathbb{R}^{d,1}$. Therefore, the discussion of symmetries under $\ISO\rbra{d,1}$ in Minkowski space $\omathbb{R}^{d,1}$ is equivalent to considering the space $\quot{\ISO\rbra{d,1}}{\SO\rbra{d,1}}$.
\end{exm}

The space $\quot{\ISO\rbra{d,1}}{\SO\rbra{d,1}}$ examined in the above example is well-suited for analyzing facts that hold up to the action of the Lorentz group $\SO\rbra{d,1}$, making it highly compatible with relativistic theories. This suggests that, not limited to the isometry group $\ISO\rbra{d,1}$, any Lie group that contains $\SO\rbra{d,1}$ as a closed subgroup can yield a coset space that is likewise suitable for relativistic considerations. The spaces obtained in this manner naturally depart from Minkowski space $\omathbb{R}^{d,1}$. Next, let us examine such examples.

\begin{exm}[(Anti-)de Sitter space]
Examples of Lie groups containing $\SO\rbra{d,1}$ as a closed subgroup include $\SO\rbra{d+1,1}$ and $\SO\rbra{d,2}$. Using these, the space defined by $\quot{\SO\rbra{d+1,1}}{\SO\rbra{d,1}}$ is called de Sitter space, and the space defined by $\quot{\SO\rbra{d,2}}{\SO\rbra{d,1}}$ is called anti-de Sitter space.
\end{exm}

The Klein geometries examined in the preceding two examples---namely Minkowski space and (anti-)de Sitter space---possess a notable property in comparison to general Klein geometries. Let us first define this property.

\begin{dfn}
Suppose that a Klein geometry $\rbra{X,G}$ has $H$ as its isotropy subgroup. It is said to be \textbf{reductive} if the Lie algebra associated with $G$ can be decomposed into a direct sum of $H$-modules, or $\Ad_{H}$-invariant vector spaces. When writing this direct sum decomposition as $\mathfrak{g}=\mathfrak{h}\oplus\mathfrak{m}$, $\mathfrak{h}$ is the Lie algebra associated with $H$, and $\mathfrak{m}$ is an $H$-module that is not necessarily a Lie algebra.
\end{dfn}

In the above definition, a subspace $\mathfrak{m}$ of $\mathfrak{g}$ is said to be an $\Ad_{H}$-invariant vector space if $\Ad_{H}{\mathfrak{m}}\subset\mathfrak{m}$ holds. That is, it means that $\mathfrak{m}$ is closed under the adjoint action of $H\subset G$. This can also be rephrased using the infinitesimal version of $\Ad_{H}$. Specifically, by considering $\Ad_{H}{\mathfrak{m}}$ near the identity element of $H$ (and thus of $G$), one can consider the vector space $\sbra{\mathfrak{h},\mathfrak{m}}$, in which case the condition that $\mathfrak{m}$ is an $\Ad_{H}$-invariant vector space induces $\sbra{\mathfrak{h},\mathfrak{m}}\subset\mathfrak{m}$.

When a Lie algebra homomorphism $I\colon\mathfrak{g}\rightarrow\mathfrak{g}$ is an involution, $I$ has eigenvalues $\pm1$. Denoting the eigenspace corresponding to the eigenvalue $-1$ by $\mathfrak{m}$, for any $X,Y\in\mathfrak{m}$, one has
\begin{align}
I\rbra{\sbra{X,Y}}=\sbra{I\rbra{X},I\rbra{Y}}=\sbra{-X,-Y}=\sbra{X,Y},
\end{align}
which impies that $\sbra{X,Y}$ belongs to the eigenspace corresponding to the eigenvalue $+1$ of $I$. From this fact and the direct sum decomposition (as vector spaces) $\mathfrak{g}=\mathfrak{h}\oplus\mathfrak{m}$, it follows that $\sbra{X,Y}$ belongs to $\mathfrak{h}$. This means that $\sbra{\mathfrak{m},\mathfrak{m}}\subset\mathfrak{h}$. Considering this, in the direct sum decomposition into $\Ad_{H}$-invariant vector spaces $\mathfrak{g}=\mathfrak{h}\oplus\mathfrak{m}$, if $\mathfrak{m}$ coincides with the eigenspace of eigenvalue $-1$ (or $\mathfrak{h}$ coincides with the eigenspace of eigenvalue $+1$), a reductive Klein geometry possessing such a decomposition satisfies $\sbra{\mathfrak{m},\mathfrak{m}}\subset\mathfrak{h}$ in addition to $\sbra{\mathfrak{h},\mathfrak{h}}\subset\mathfrak{h}$ and $\sbra{\mathfrak{h},\mathfrak{m}}\subset\mathfrak{m}$.

\begin{exm}[Maximally symmetric spaces]
The spaces in the previous examples---namely Minkowski space $\quot{\ISO\rbra{d,1}}{\SO\rbra{d,1}}$, de Sitter space $\quot{\SO\rbra{d+1,1}}{\SO\rbra{d,1}}$, and anti-de Sitter space $\quot{\SO\rbra{d,2}}{\SO\rbra{d,1}}$---are collectively called \textbf{maximally symmetric spaces}, and all of them constitute reductive Klein geometries. That is to say, $\quot{\iso\rbra{d,1}}{\so\rbra{d,1}}$, $\quot{\so\rbra{d+1,1}}{\so\rbra{d,1}}$, and $\quot{\so\rbra{d,2}}{\so\rbra{d,1}}$ from $\Ad_{\SO\rbra{d,1}}$-invariant vector spaces, satisfying inclusion relations for the Lie brackets analogous to those in the general framework discussed above. We will revisit this point in a later section when discussing gravity.
\end{exm}

\vspace{\baselineskip}

This concludes our discussion of Klein geometry. Next, let us introduce deformations to this framework.

\subsection{Cartan geometry}
While Klein geometry being a framework in which the relation between a space and the group acting upon it clearly understood, the spaces one wishes to consider are not always of this form (i.e., expressible as homogeneous spaces or coset spaces)\footnote{For instance, when describing spacetime distorted by the presence of celestial boodies, Klein geometries such as Minkowski space or (anti-)de Sitter space prove insufficient.}. Thus, as an approach to such more general spaces or geometries, we introduce a geometry that extends the notion of Klein geometry. By employing this geometry, one can naturally formulate gravitational theories not only on Minkowski spacetime but also on spacetimes with non-vanishing cosmological constant\footnote{A concrete gravitational theory based on this perspective is introduced in Section~\ref{sec:1FSoLGfSUC}.}.

\subsubsection{Cartan geometry and its connection}
Let us define Cartan geometry as a geometry that locally resembles a Klein geometry. Before stating the precise definition, we outline the overview; a Cartan geometry consists of a principal $H$-bundle $\mathscr{G}\rightarrow M$ over a manifold $M$, associated with a closed subgroup $H$ of a Lie group $G$, equipped with a $\mathfrak{g}$-valued $1$-form on $\mathscr{G}$ called a Cartan connection that satisfies certain properties (where $\mathfrak{g}$ is the Lie algebra associated with $G$). The properties that a Cartan connection must consist of two parts: the relationship between the tangent space and $\mathfrak{g}$, and the transformation law under the action of $H$. Let us define Cartan geometry in more detail regarding these aspects.

\begin{dfn}[Cartan geometry]
Let $\rbra{X,G}$ be a Klein geometry, $H$ be the isotropy subgroup of $G$ at a point in $X$, and $\mathfrak{g}$ and $\mathfrak{h}$ be the Lie algebras associated with $G$ and $H$, respectively. Furthermore, let $M$ be a smooth manifold with the same dimention as $X$. When there exists a $\mathfrak{g}$-valued $1$-form $\omega\in\Omega^{1}\rbra{\mathscr{G},\mathfrak{g}}$ on a principal $H$-bundle $\mathscr{G}\rightarrow M$ over $M$ satisfying the following properties:
\begin{itemize}
\item
At each $p\in\mathscr{G}$, $\omega\vert_{p}\colon T_{p}\mathscr{G}\rightarrow\mathfrak{g}$ is a linear isomorphism.
\item
For any $h\in H$, $\rbra{\triangleleft h}^{\ast}\omega=\Ad_{h}^{-1}{\omega}$ holds.
\item
For the fundamental vector field\footnote{For an element $X\in\mathfrak{h}$ of the Lie algebra, the vector field $X^{\mathscr{G}}\in\mathfrak{X}\rbra{\mathscr{G}}$ induced on $\mathscr{G}$ is defined at each point $p\in\mathscr{G}$ by
\begin{align}
X^{\mathscr{G}}\vert_{p}\coloneqq\frac{d}{dt}\biggr\vert_{t=0}p\triangleleft\exp{\rbra{tX}}.
\end{align}
This $X^{\mathscr{G}}$ is called the fundamental vector field on $\mathscr{G}$ associated with $X$. Here, $\exp\colon\mathfrak{h}\rightarrow H$ is the exponential map.} $X^{\mathscr{G}}\in\mathfrak{X}\rbra{\mathscr{G}}$ for any $X\in\mathfrak{h}$, $\omega\vert_{p}\rbra{X^{\mathscr{G}}}=X$ holds at each $p\in\mathscr{G}$.
\end{itemize}
the tuple $\rbra{M,\mathscr{G},\omega}$ is called a \textbf{Cartan geometry}. In this context, the Klein geometry $\rbra{X,G}$ is called the \textbf{model geometry} of the Cartan geometry $\rbra{M,\mathscr{G},\omega}$.
\end{dfn}

While a Cartan connection is defined similarly to an Ehresmann connection, there is a ctucial difference. This lies in the Lie algebra in which the connection takes its values: whereas an Ehresmann connection takes values in the Lie algebra associated with the structure group of the principal bundle (which is diffeomorphic to each fiber), a Cartan connection takes values in the Lie algebra associated with a group strictly lager than the structure group. As we shall see later, this is the reason why a Cartan geometry can be locally viewed as a Klein geometry. For the time being, let us see through the following example that Cartan geometry serves as an extension of Klein geometry.

\begin{exm}[Klein geometry]
Let a Klein geometry $\rbra{X,G}$ have $H\subset G$ as the isotropy subgroup at a suitable point. Then $X$ is diffeomorphic to $\quot{G}{H}$ and naturally inherits the structure of a principal $H$-bundle $G\rightarrow X$. Namely, this structure is given by the commutative diagram:
\begin{center}
\begin{tikzcd}[row sep=1cm,column sep=2.5cm]
&G\arrow[dl,"\text{natural projection}"']\arrow[d,"\text{principal $H$ bundle}"]\\
\quot{G}{H}\arrow[r,"\text{diffeomorphism}"']&X
\end{tikzcd}
\end{center}
Meanwhile, on the total space of the Lie group $G$, there exists a $\mathfrak{g}$-valued $1$-form $\omega_{G}$ on $G$, known as the Maurer-Cartan form (where $\mathfrak{g}$ is the Lie algebra associated with $G$). Using left translation
\begin{align}
\begin{array}{r@{\,\,}c@{\,\,}c@{\,\,}c@{\,\,}}
L\colon&G\times G&\longrightarrow&G\\
&\rotatebox{90}{$\in$}&&\rotatebox{90}{$\in$}\\
&\rbra{g,g^{\prime}}&\longmapsto&L_{g}\rbra{g^{\prime}}\coloneqq gg^{\prime}
\end{array},
\end{align}
it is defined at each point $g\in G$ for any tangent vector $v\in T_{g}G$ by $\omega_{G}\vert_{g}\rbra{v}\coloneqq\rbra{L_{g^{-1}}}_{\ast,g}\rbra{v}$\footnote{$\omega_{G}\vert_{g}\rbra{v}$ is a tangent vector at the identity element of $G$, which is identified with an element of $\mathfrak{g}$.}. This Maurer-Cartan form $\omega_{G}$ can be regarded as a Cartan connection on the principal $H$-bundle $G\rightarrow X$. To verify this, three conditions must be satisfied, which we check in order. First, as to whether $\omega_{G}\vert_{g}$ is a linear isomorphism from $T_{g}G$ to $\mathfrak{g}$ at each $g\in G$, this is guaranteed by the fact that $T_{g}G$ is linearly isomorphic to the tangent space at the identity element of $G$, which in turn is linearly isomorphic to $\mathfrak{g}$. Next, since the map $\triangleleft h\colon G\rightarrow G$ correspods to right translation for any $h\in H$:
\begin{align}
\begin{array}{r@{\,\,}c@{\,\,}c@{\,\,}c@{\,\,}}
R\colon&G\times G&\longrightarrow&G\\
&\rotatebox{90}{$\in$}&&\rotatebox{90}{$\in$}\\
&\rbra{g,g^{\prime}}&\longmapsto&R_{g}\rbra{g^{\prime}}\coloneqq g^{\prime}g
\end{array},
\end{align}
it follows that $\rbra{\triangleleft h}^{\ast}\omega_{G}\vert_{g}\rbra{v}=\omega_{G}\vert_{g}\rbra{\rbra{R_{g}}_{\ast,g}v}=\rbra{L_{g^{-1}}\circ R_{g}}_{\ast,g}\rbra{v}$, which implies that $\rbra{\triangleleft h}^{\ast}\omega_{G}=\Ad_{h}^{-1}{\omega_{G}}$ also holds. Finally, for the fundamental vector field $X^{G}$ associated with an arbitrary element $X$ of the Lie algebra $\mathfrak{h}$ of $H$, we have $\omega_{G}\vert_{g}\rbra{X^{G}}=\rbra{L_{g^{-1}}}_{\ast,g}\rbra{X^{G}\vert_{g}}=X$ at each point, so the last condition is satisfied as well. From the above, the triple $\rbra{X,G,\omega_{G}}$ inheed constitutes a Cartan geometry.
\end{exm}

Let us also examine a more non-trivial and physically relevant example of Cartan geometry. It provides a spacetime picture in which a maximally symmetric space, viewed as a Klein geometry, serves as the vacuum.

\begin{exm}[Lorentz geometry]
\label{exm:LGiCG}
Lorentzian geometry (or more generally, pseudo-Riemannian geometry) serves as an example of Cartan geometry. Since Lorentzian geometry essentially deals mainly with objects invariant under the local action of the Lorentz group, this corresponds to choosing the Lorentz group as the closed subgroup in Cartan geometry. For a Lorentzian manifold $\rbra{M,g}$ equipped with a Lorentzian metric $g$ on a manifold $M$, let us prepare the orthonormal frame bundle of $M$ with respect to $g$. Taking the signature to be $\rbra{d,1}$, the structure group of this principal bundle is the orthogonal group $\O\rbra{d,1}$. While we could proceed with this choice, we shall consider a frame bundle $\mathscr{F}\rightarrow M$ whose structure group is reduced to the Lorentz group $\SO\rbra{d,1}$ to better reflect the physical context.

Under the above setup, let us consider the isometry group $\ISO\rbra{d,1}$ having $\SO\rbra{d,1}$ as a closed subgroup, and prepare a $1$-form $\mathcal{A}$ on $\mathscr{F}$ taking values in the Lie algebra $\iso\rbra{d,1}$ associated with $\ISO\rbra{d,1}$. Considering the direct sum decomposition as vector spaces
\begin{align}
\iso\rbra{d,1}\cong\so\rbra{d,1}\oplus\quot{\iso\rbra{d,1}}{\so\rbra{d,1}}
\end{align}
$\mathcal{A}$ can viewed as a direct sum $\rbra{\omega,e}$ of a $1$-form $\omega$ taking values in $\so\rbra{d,1}$ and a $1$-form $e$ taking values in $\quot{\iso\rbra{d,1}}{\so\rbra{d,1}}$. When we take the spin connection (the Ehresmann connection on the frame bundle $\mathscr{F}\rightarrow M$) and the vielbein as this $\omega$ and $e$, respectively, then $\mathcal{A}=\rbra{\omega,e}$ becomes a Cartan connection on $\mathscr{F}$. Hence, $\rbra{M,\mathscr{F},\mathcal{A}}$ constitutes a Cartan geometry. Now, as for what its model geometry is, it is the pair $\rbra{\omathbb{R}^{d,1},\ISO\rbra{d,1}}$ consisting of Minkowski space $\omathbb{R}^{d,1}$, which is diffeomorphic to $\quot{\ISO\rbra{d,1}}{\SO\rbra{d,1}}$, and the isometry group $\ISO\rbra{d,1}$.

Taking de Sitter space or anti-de Sitter space as model geometries---namely $\rbra*{\quot{\SO\rbra{d+1,1}}{\SO\rbra{d,1}},\SO\rbra{d+1,1}}$ or $\rbra*{\quot{\SO\rbra{d,2}}{\SO\rbra{d,1}},\SO\rbra{d,2}}$---one can similarly consider a Cartan geometry whose Cartan connection is an $\so\rbra{d+1,1}$-valued or $\so\rbra{d,2}$-valued $1$-form on the $\SO\rbra{d,1}$-frame bundle $\mathscr{F}\rightarrow M$. As in the case of Minkowski space, the Cartan connection in these cases can also be expressed as the direct sum of a spin connection and a vielbein. Therefore, Minkowski space, de Sitter space, and anti-de Sitter space each serve as a model geometry for the Cartan geometry $\rbra{M,\mathscr{F},\mathcal{A}}$ by appropriately choosing (the target algebra of) the Cartan connection $\CMcal{A}$.
\end{exm}

Once a connection is defined, its curvature is introduced naturally. Because this curvature plays a crucial role in describing gravity on maximally symmetric spaces, we shall devote a separate section to its discussion.

\subsubsection{Curvature of a Cartan connection}
Just as the curvature of an Ehresmann connection is defined and plays various important roles in mathematics and physics, a useful notion of curvature can likewise be defined for a Cartan connection.

\begin{dfn}[Curvature of a Cartan connection]
In Cartan geometry $\rbra{M,\mathscr{G},\omega}$ with a Klein geometry $\rbra{X,G}$ as its model geometry, the curvature of the Cartan connection $\omega$ is defined as the $\mathfrak{g}$-valued $2$-form
\begin{align}
\varOmega\coloneqq d\omega+\frac{1}{2}\wedgespair{\omega}{\omega}.
\end{align}
Here, $\wedgespair{-}{-}$ combines the Lie bracket $\sbra{-,-}$ on the Lie algebra $\mathfrak{g}$ with the exterior product $\wedge$.
\end{dfn}

Although the curvature of a Cartan connection appears superficially identical to that of an Ehresmann connection, it is important to note that they take values in different Lie algebras. Because the curvature of a Cartan connection takes values in the Lie algebra of a group strictly larger than the structure group, it can encode richer information compared to the curvature of an Ehresmann connection, which takes values only in the Lie algebra associated with the structure group. Let us examine examples of this below.

\begin{exm}[Lorentz geometry]
\label{exm:CoCCoLG}
Let us consider the Cartan geometry $\rbra{M,\mathscr{F},\mathcal{A}}$ corresponding to Lorentzian geometry in Example~\ref{exm:LGiCG}. Calculating the curvature $\mathscr{F}$ of the Cartan connection $\mathcal{A}=\rbra{\omega,e}$, we obtain
\begin{align}
\mathcal{F}\coloneqq d\mathcal{A}+\frac{1}{2}\wedgespair{\mathcal{A}}{\mathcal{A}}=\rbra*{d\omega+\frac{1}{2}\wedgespair{\omega}{\omega}+\frac{1}{2}\wedgespair{e}{e},de+\wedgespair{\omega}{e}}.
\end{align}
From this, one sees that $\mathscr{F}$ is a combination of the curvature form $d\omega+\frac{1}{2}\wedgespair{\omega}{\omega}$ and the torsion form $de+\wedgespair{\omega}{e}$ in Lorentzian geometry. The term $\frac{1}{2}\wedgespair{e}{e}$ added to the curvature form vanishes when Minkowski space is chosen as the model geometry, and physically, it is determined by the cosmological constant. In this way, understanding Lorentzian geometry as a Cartan geometry allows the curvature in Cartan geometry to be captured as a unified concept encompassing both curvature and torsion in Lorentzian geometry. Consequently, the vanishing of the curvature $\mathscr{F}$ of the Cartan connection $\mathcal{A}$ is equivalent to the simultaneous vanishing of both the Riemannian curvature and the torsion.
\end{exm}

When the curvature of a Cartan connection vanishes, such a Cartan conneciton---or the corresponding Cartan geometry---is said to be flat. The meaning of this term should be clear from the preceding example. In connection with this, let us verify the curvature in the case of a Klein geometry.

\begin{exm}[Klein geometry]
In a Klein geometry $\rbra{X,G,\omega_{G}}$ viewed as a Cartan geometry, the Maurer-Cartan form $\omega_{G}$, which serves as the Cartan connection, satisfies the structure equation
\begin{align}
d\omega_{G}+\frac{1}{2}\wedgespair{\omega_{G}}{\omega_{G}}=0.
\end{align}
Since the left-hand side of this equation is precisely the curvature of Cartan connection $\omega_{G}$, this demonstrates that a Klein geometry is a flat Cartan geometry.
\end{exm}

In the examples of Lorentzian geometry such as Example~\ref{exm:LGiCG} and Example~\ref{exm:CoCCoLG}, we saw the relationship between the connections, curvatures, and torsion in conventional geometry and those in Cartan geometry by splitting the Cartan connection and its curvature into two parts according to the space in which they take values. Such a perspective of decomposing the Lie algebra (as a vector space) can also be applied to general Cartan geometries. Namely, in a Cartan geometry $\rbra{M,\mathscr{G},\omega}$ with a Klein geometry $\rbra*{\quot{G}{H},G}$ as its model geometry, the Lie algebra in which the Cartan connection $\omega$ and its curvature $\varOmega$ take values is the Lie algebra $\mathfrak{g}$ associated with $G$. As a vector space, this can be decomposed into a direct sum of the Lie algebra $\mathfrak{h}$ associated with $H$ and the quotient space $\quot{\mathfrak{g}}{\mathfrak{h}}$:
\begin{align}
\mathfrak{g}\cong\mathfrak{h}\oplus\quot{\mathfrak{g}}{\mathfrak{h}}.
\end{align}
Through this decomposition, both $\omega$ and $\varOmega$ can be divided into an $\mathfrak{h}$-component and a $\quot{\mathfrak{g}}{\mathfrak{h}}$-component. Since $\mathscr{G}$ is a principal $H$-bundle, the $\mathfrak{h}$-component is analogous to that in the Ehresmann case; indeed, in Example~\ref{exm:LGiCG} and Example~\ref{exm:CoCCoLG}, this component corresponded to the Ehresmann connection and its curvature on the frame bundle. On the other hand, the $\quot{\mathfrak{g}}{\mathfrak{h}}$-component vividly reflects the difference from the Ehresmann case. The $\quot{\mathfrak{g}}{\mathfrak{h}}$-component of the Cartan connection $\omega$, given by $\sigma\coloneqq p_{\text{quot}}\rbra{\omega}\in\Omega^{1}\rbra*{\mathscr{G},\quot{\mathfrak{g}}{\mathfrak{h}}}$ via the natural projection $p_{\text{quot}}\colon\mathfrak{g}\rightarrow\quot{\mathfrak{g}}{\mathfrak{h}}$, is called the \textbf{solder form}. The pullback of this form $e\coloneqq s^{\ast}\sigma\in\Omega^{1}\rbra*{U,\quot{\mathfrak{g}}{\mathfrak{h}}}$ by a section $s\in\Gamma\rbra{U,\mathscr{G}}$ over an open set $U\subset M$ of the base space (a local section on $M$) is the field known in physics (Lorentzian geometry) as the vielbein. On the other hand, the pullback of the $\mathfrak{h}$-component of the Cartan connection by a local section $s$ is referred to in physics (Lorentzian geometry) as the spin connection. This corresponds to the pullback of Ehresmann connection by a local section being called the Yang-Mills field or the Yang-Mills-Utiyama field.

\vspace{\baselineskip}

For a more comprehensive treatment of Cartan geometry, we refer the reader to the literature; here, we turn our attention to the formulation of gravity within this geometric framework.

\section{$1$-form symmetry of linearized gravity with small cosmological constant}
\label{sec:1FSoLGfSUC}
In this section, we add a cosmological constant term to the linearized gravity theory described by the vielbein and the spin connection expanded up to first order in $\sqrt{G}$, and carry out our discussion by expanding the theory in a power series with respect to the cosmological constant $\Lambda$. Through this analysis, we shall see that under certain circumstances, a $1$-form symmetry exists even in the presence of a cosmological term.

In what follows, we first develop the background geometry of spacetimes with a cosmological term, following MacDowell and Mansouri~\cite{MacDowell-Mansouri}, as well as Wise~\cite{Wise2}. Here, we review how Cartan geometry, introduced in the previous section, describes gravity. Subsequently, we examine the properties of the gravitational fields in regimes where gravitational effects are relatively small.

Unless othewise stated, we assume the spacetime dimension to be $3+1$ throughout this section.

\subsection{Description via the MacDowell-Mansouri formalism}
To describe (classical) gravitational theories with a cosmological constant, it is convenient to employ the MacDowell-Mansouri formalism, which has Cartan geometry as its underlying geometric framework. Let us first briefly review this formulation. The original paper by MacDowell and Mansouri is~\cite{MacDowell-Mansouri}, and a remarkably clear exposition together with the geometric picture is given by Wise~\cite{Wise2}.

\subsubsection{Geometry describing spacetime}
Let us consider describing a gravitational theory on a $\rbra{3+1}$-dimensional spacetime within the MacDowell-Mansouri formalism. As is well known, on  general $\rbra{d+1}$-dimensional Lorentzian manifold, there exist three maximally symmetric spaces (spaces of constant curvature) characterized by the sign (or vanishing) of the cosmological constant $\Lambda$:
\begin{center}
{\setlength{\tabcolsep}{30pt}
\begin{tabular}{ccc}
$\Lambda<0$&$\Lambda=0$&$\Lambda>0$\\[10pt]
$\quot{\SO\rbra{d,2}}{\SO\rbra{d,1}}$&$\quot{\ISO\rbra{d,1}}{\SO\rbra{d,1}}$&$\quot{\SO\rbra{d+1,1}}{\SO\rbra{d,1}}$\\[10pt]
anti-de Sitter&Minkowski&de Sitter
\end{tabular}}
\end{center}
Here, $\SO\rbra{s,t}$ denotes the group that leaves invariant a metric with $s$ positive eigenvalues and $t$ negative eigenvalues, and $\ISO\rbra{s,t}$ refers to the semidirect product $\SO\rbra{s,t}\ltimes\omathbb{R}^{s,t}$ of $\SO\rbra{s,t}$ and $\omathbb{R}^{s,t}$.

All of these spacetimes are obtained by identifying a Lie group modulo one of its subgroups. Spacetime geometry that can be formulated in terms of groups in this way can be described within the framework of Klein geometry. Therefore, a reference spacetime (in the absence of matter or other gauge fields) can be treated as a homogeneous space using the language of Klein geometry. This generalizes the case with a vanishing cosmological constant, where Minkowski spacetime served as the reference background. Just as Minkowski spacetime curves into a general Lorentzian spacetime in the presence of matter, when a homogeneous space as a spacetime undergoes analogous deformation, it is naturally described by Cartan geometry. That is, this suggests that Cartan geometry---rather than Lorentzian (or pseudo-Riemannian) geometry---is the more natural framework for describing spacetimes with a cosmological constant. Consequently, the MacDowell-Mansouri formulation of gravity discussed below will be developed in the language of Cartan geometry.

When addressing gravity and spacetime in the language of Cartan geometry, this framework naturally offers the following perspective on spacetime deformation (i.e., gravity), there are two distinct types of effects that contribute to how spacetime is deformed, each originating from a different parameter. The first is the deformation of the local homogeneous space---the local picture of spacetime represented in Cartan geometry---away from Minkowski space; this defoemation is induced by the cosmological term, with its magnitude determined by the cosmological constant. The second is the degree to which the spacetime described by Cartan geometry deviates from this homogeneous space, this is generated by matter energy-momentum, with its magnitude determined by the gravitational constant. Thus, what is broadly called spacetime deformation actually consists of two natural modes of deformation within Cartan geometry. Consequently, when analyzing the gravitational field using Cartan geometry, it is crucial to remain mindful of these two distinct parameters: the cosmological constant and the gravitational constant. This distinction plays an essential role, particularly when discussing the properties of the gravitational field in spacetimes with small deformations.

\vspace{\baselineskip}

These homogeneous spaces---namely Minkowski space and (anti-)de Sitter space, collectively referred to as maximally symmetric spaces---possess a special property that does not necessarily hold for general homogeneous spaces. Namely, they are reductive; in other words, setting $\mathfrak{g}=\iso\rbra{d,1},\so\rbra{d+1,1},\so\rbra{d,2}$, the vector space decomposition
\begin{align}
\mathfrak{g}\cong\so\rbra{d,1}\oplus\quot{\mathfrak{g}}{\so\rbra{d,1}}
\end{align}
also carries the meaning of a decomposition as Lie algebras. However, this does not imply that $\quot{\mathfrak{g}}{\so\rbra{d,1}}$ is a Lie subalgebra, but rather that $\quot{\mathfrak{g}}{\so\rbra{d,1}}$ can be regarded as an $\Ad_{\SO\rbra{d,1}}$-invariant vector space. Furthermore, since $\so\rbra{d,1}$ is naturally an $\Ad_{\SO\rbra{d,1}}$-invariant vector space, the reductivity of these maximally symmetric spaces means that the above decomposition corresponds to a decomposition of $\SO\rbra{d,1}$ representations under the adjoint action.

From the reasoning above, we see that the bracket relations
\begin{align}
\sbra{\so\rbra{d,1},\so\rbra{d,1}}\subset\so\rbra{d,1},\qquad\sbra*{\so\rbra{d,1},\quot{\mathfrak{g}}{\so\rbra{d,1}}}\subset\quot{\mathfrak{g}}{\so\rbra{d,1}},\qquad\sbra*{\quot{\mathfrak{g}}{\so\rbra{d,1}},\quot{\mathfrak{g}}{\so\rbra{d,1}}}\subset\so\rbra{d,1}
\end{align}
hold as Lie algebra inclusions. For the basis-dependent expression of these relations, see Eq.~\eqref{eq:LB} below. This will later be used to gauge the effects of the cosmological constant in the discussion of higher-form symmetries.

\subsubsection{MacDowell-Mansouri formalism}
The known fundamental interactions other than gravity---namely electromagnetism, the strong force, and the weak force---have their mathematical foundation in the geometry of Ehresmann connections (which we shall refer to simply as Ehresmann geometry). This is arguably the most suitable language available today.

What, then, is the appropriate language for describing gravity? Here, we take it to be Cartan geometry. The relation between gauge theories based on Ehresmann geometry and those (i.e., gravitational theories) based on Cartan geometry is summarized schematically below. The interaction between gravity and other gauge or matter fields takes place via the Hodge star operator.

The right side of the figure below illustrates the formulation of interactions other than gravity, built upon the mathematical foundation of Ehresmann geometry, where the gauge field is obtained as the pullback of an Ehresmann connection to spacetime. Conversely, the left side depicts the formulation of gravity rooted in Cartan geometry, where the pullbacks of the Cartan connection give rise to the spin connection and the vielbein. The bridge connecting these two formulations is the Hodge star operator. By constructing the Hodge operator from the vielbein arising in Cartan geometry and applying it to other fields, such as gauge fields, within the action functional, the interaction between the gravitational field and other fields is naturally realized.

\begin{center}
\begin{tikzcd}[row sep=5cm,column sep=2.5cm]
\text{Cartan geometry}\arrow[d,"\text{MacDowell-Mansouri formalism}"']&\text{Ehresmann geometry}\arrow[d,""'{name=YMU},"\text{Yang-Mills-Utiyama formalism}"]\\
\text{gravitational theory}&\text{gauge theory (other than gravity)}\arrow[from=1-1,to=YMU,out=-80,in=135,"\text{Hodge star operator}"description]
\end{tikzcd}
\end{center}

In what follows, we briefly discuss the leftmost arrow in the diagram above as preparation for our subsequent analysis of symmetries.

\paragraph{Gauge theory on Cartan geometry}
Let $M$ be a $\rbra{d+1}$-dimensional manifold representing spacetime (henceforth referred to simply as spacetime), and let $G$ be a Lie group containing the Lorentz group $H=\SO\rbra{d,1}$ as a closed subgroup. For instance, if one wishes to consider a spacetime based on Minkowski space ($\Lambda=0$), one sets $G=\ISO\rbra{d,1}=\SO\rbra{d,1}\ltimes\omathbb{R}^{d,1}$; for a spacetime based on de Sitter space ($\Lambda>0$), $G=\SO\rbra{d+1,1}$; and for a spacetime based on anti-de Sitter space ($\Lambda<0$), $G=\SO\rbra{d,2}$.

The geometry serving as the model for spacetime is the Klein geometry $\rbra*{\quot{G}{H},G}$, and if matter energy-momentum is present, spacetime becomes a Cartan geometry deformed away from this baseline model. As in standard general relativity, the manner of this deformation is governed by the Einstein equations. The deformation in Cartan geometry away from the background Klein geometry is expressed precisely in terms of the curvature of the Cartan connection. Therefore, this curvature---or the Cartan connection that determines it---represents gravity, and the resulting equations of motion are equivalent to the Einstein equations. Note, however, that since the Cartan connection and its curvature are differential forms defined on a principal bundle over spacetime rather than on spacetime itself, it is necessary to select an appropriate section (gauge) to pull then back onto spacetime. This point requires caution just as in standard gauge theories.

Let us consider a Cartan geometry $\rbra{M,\mathscr{G},\mathcal{A}}$ with model geometry given by the Klein geometry $\rbra*{\quot{G}{H},G}$. Here, $\mathscr{G}$ is a principal $H$-bundle over the spacetime $M$, and $\mathcal{A}\in\Omega^{1}\rbra{\mathscr{G},\mathfrak{g}}$ is a Cartan connection on it. As usual, $\mathfrak{g}$ denotes the Lie algebra associated with $G$\footnote{If $\Lambda=0$, then $\mathfrak{g}=\iso\rbra{d,1}=\so\rbra{d,1}\oplus_{\mathsf{S}}\omathbb{R}^{d,1}$ (semidirect sum); if$\Lambda>0$, then $\mathfrak{g}=\so\rbra{d+1,1}$; and if $\Lambda<0$, then $\mathfrak{g}=\so\rbra{d,2}$.}, and the Lie algebra associated with $H$ is written as $\mathfrak{h}=\so\rbra{d,1}$. Let $\text{\textomega}\in\Omega^{1}\rbra{\mathscr{G},\mathfrak{h}}$ be the $\mathfrak{h}$-component of the Cartan connection $\mathcal{A}$, which serves as an Ehresmann connection, and let $\text{\textepsilon}\in\Omega^{1}\rbra*{\mathscr{G},\quot{\mathfrak{g}}{\mathfrak{h}}}$ be the $\quot{\mathfrak{g}}{\mathfrak{h}}$-component, serving as the solder form. We express this decomposition as $\mathcal{A}=\rbra{\text{\textomega},\text{\textepsilon}}$ or $\mathcal{A}=\text{\textomega}+\text{\textepsilon}$\footnote{In this decomposition, a constant derived from the cosmological constant is sometimes multiplied in front of the $\quot{\mathfrak{g}}{\mathfrak{h}}$-component; however, in this work, we absorb this constant into the structure constants of the Lie algebra characterized by the cosmological constant.}. If we denote the pullbacks of $\mathcal{A}$ and its components by a suitable section $s\in\Gamma\rbra{U,\mathscr{G}}$ over an open set $U\subset M$ as $A\coloneqq s^{\ast}\mathcal{A}\in\Omega^{1}\rbra{U,\mathfrak{g}}$, $\omega\coloneqq s^{\ast}\text{\textomega}\in\Omega^{1}\rbra{U,\mathfrak{h}}$, and $e\coloneqq s^{\ast}\text{\textepsilon}\in\Omega^{1}\rbra*{U,\quot{\mathfrak{g}}{\mathfrak{h}}}$, then $\omega$ and $e$ become the spin connection and the vielbein (on $U$), respectively, which serve as the fields governing gravity on spacetime\footnote{Whether a global section of $\mathscr{G}$ over $M$ exists is equivalent to whether $\mathscr{G}$ is bundle-isomorphic to the trivial bundle $M\times H$ over M. That is, if $\mathscr{G}$ is trivial, a global section exists, and vice versa.}. Therefore, as a theory of the gravitational field, it suffices to consider a theory formulated in terms of $A$, or equivalently, $\omega$ and $e$.

Let $\mathcal{F}\coloneqq d\mathcal{A}+\frac{1}{2}\wedgespair{\mathcal{A}}{\mathcal{A}}\in\Omega^{2}\rbra{\mathscr{G},\mathfrak{g}}$ be the curvature of the Cartan connection $\mathcal{A}$, and denote the $\mathfrak{h}$-component and $\quot{\mathfrak{g}}{\mathfrak{h}}$-component of its pullback $F\coloneqq s^{\ast}\mathcal{F}\in\Omega^{2}\rbra{U,\mathfrak{g}}$ via a local section $s$ by $\widehat{F}\in\Omega^{2}\rbra{U,\mathfrak{h}}$ and $T\in\Omega^{2}\rbra*{U,\quot{\mathfrak{g}}{\mathfrak{h}}}$, respectively. The former represents the Riemann curvature of spacetime modified by the cosmological constant, while the latter represents the torsion of spacetime. Expressing $F$ in terms of $A=\omega+e$, we have
\begin{align}
F=dA+\frac{1}{2}\wedgespair{A}{A}={}\underbrace{\underbrace{d\omega+\frac{1}{2}\wedgespair{\omega}{\omega}}_{=R}{}+\wedgespair{e}{e}}_{=\widehat{F}}{}+{}\underbrace{de+\wedgespair{\omega}{e}}_{=T},
\end{align}
which confirms that $F$ indeed encodes information regarding the Riemann curvature and torsion of spacetime. Here, the term designated as $R$ represents the Riemann curvature $2$-form on spacetime.

\paragraph{Basis-dependent representation}
Let $\cbra{\mathsf{M}_{ab}}_{a,b=0}^{d}$ be a basis for the Lie algebra $\mathfrak{h}=\so\rbra{d,1}$, and let $\cbra{\mathsf{P}_{a}}_{a=0}^{d}$ be a basis for $\quot{\mathfrak{g}}{\mathfrak{h}}$. The Lie brackets among these basis generators are given by
\begin{align}
\label{eq:LB}
\sbra{\mathsf{M}_{ab},\mathsf{M}_{cd}}=\rbra*{\eta_{bc}\delta^{e}_{a}\delta^{f}_{d}+\eta_{ad}\delta^{e}_{b}\delta^{f}_{c}-\eta_{ac}\delta^{e}_{b}\delta^{f}_{d}-\eta_{bd}\delta^{e}_{a}\delta^{f}_{c}}\mathsf{M}_{ef},\quad\sbra{\mathsf{M}_{ab},\mathsf{P}_{c}}=\rbra*{\eta_{bc}\delta^{d}_{a}-\eta_{ac}\delta^{d}_{b}}\mathsf{P}_{d},\quad\sbra{\mathsf{P}_{a},\mathsf{P}_{b}}=-\frac{\lambda}{l^{2}}\mathsf{M}_{ab}.\notag\\
\end{align}
Here, $\lambda$ and $l$ are numbers determined according to $\mathfrak{g}$ (or the cosmological constant): if $\mathfrak{g}=\so\rbra{d+1,1}$ ($\Lambda>0$), then $\lambda=1$; if $\mathfrak{g}=\iso\rbra{d,1}$ ($\Lambda=0$), then $\lambda=0$; and if $\mathfrak{g}=\so\rbra{d,2}$ ($\Lambda<0$), then $\lambda=-1$. Futhermore, the positive real constant $l$ satisfies
\begin{align}
\frac{1}{l^{2}}=\lambda\frac{\Lambda}{3}.
\end{align}
Using these relations, the curvature $F$, expressed in terms of $\omega=\frac{1}{2}\omega^{ab}\otimes\mathsf{M}_{ab}$ and $e=e^{a}\otimes\mathsf{P}_{a}$, is written as
\begin{align}
F=\frac{1}{2}\rbra*{d\omega^{ab}+\eta_{cd}\omega^{ac}\wedge\omega^{db}-\frac{\lambda}{l^{2}}e^{a}\wedge e^{b}}\otimes\mathsf{M}_{ab}+\rbra*{de^{a}+\eta_{bc}\omega^{ab}\wedge e^{c}}\otimes\mathsf{P}_{a}.
\end{align}

In the commutation relations among the basis generators, the relation between $\mathsf{P}_{a}$ and $\mathsf{P}_{b}$ in particular can be regarded as that of an abelian Lie algebra provided $l$ is sufficiently large. This fact will play a crucial role in what follows.

\paragraph{MacDowell-Mansouri action functional}
The dynamics of the gravitational field $A$ on a $\rbra{3+1}$-dimensional spacetime $M$ (or an open set $U\subset M$) whose model geometry is de Sitter space or anti-de Sitter space is governed by the MacDowell-Mansouri action functional:
\begin{align}
\label{eq:MMAF}
S\sbra{A}=-\frac{l^{2}}{2\lambda G}\int_{M}\abra{\widehat{F},\widehat{F}}.
\end{align}
Here, $G$ is Newton's gravitational constant, and $\widehat{F}$ is the $\so\rbra{d,1}$-component of the curvature $F$ of $A$. Furthermore, $\abra{-,-}$ denotes an inner product on $\Omega^{k}\rbra{U,\mathfrak{h}}$ ($k=0,\dots,d+1$), which in the present case can be rewritten specifically as $\abra{\widehat{F},\widehat{F}}=\tr_{\mathfrak{h}}{\rbra*{\widehat{F}\wedge{\star_{\text{int}}}\widehat{F}}}$. As discussed shortly below, if one wishes to understand the gravitational dynamics including the case where Minkowski space serves as the model geometry, a slight modification to the above action is required (though this modification does not affect the equations of motion at the classical level).

Decomposing $\widehat{F}$ in the action functional into the Riemann curvature $R$ and the cosmological term $\wedgespair{e}{e}$, the MacDowell-Mansouri action functional is rewritten as
\begin{align}
S\sbra{\omega,e}=-\frac{l^{2}}{\lambda G}\int_{M}\abra{R,\wedgespair{e}{e}}+\frac{1}{2}\abra{\wedgespair{e}{e},\wedgespair{e}{e}}+\frac{1}{2}\abra{R,R},
\end{align}
where the argument $A$ of $S$ has also been split into $\omega$ and $e$. The first two terms correspond to the well-known Palatini action functional with a cosmological constant. The last term is a topological term representing the Euler number, whose presence or absence does not affect the gravitational dynamics (i.e., the equations of motion) at the classical level\footnote{Varying this term with respect to $\omega$ yields the Bianchi identity $d_{\omega}R=0$, which holds geometrically and thus does not constitute an equation of motion.}. When Minkowski space is taken as the model geometry, this last term must be subtracted before taking $\lambda=0$ ($l\to0$), thereby preventing its divergence from entering the action functional. Consequently, only the first term in the above action functional remains, reproducting the standard Palatini action functional with Minkowski space as its model geometry.

\vspace{\baselineskip}

The equations of motion derived from the MacDowell-Mansouri action functional are equivalent to the Einstein equations (with a cosmological constant).

\subsection[Linearized gravity under small deformations from Minkowski spacetime and the associated $1$-form symmetry]{\small Linearized gravity under small deformations from Minkowski spacetime and the associated $1$-form symmetry}
This section constitutes the primary contribution of the present paper. For the three Klein geometries considered thus far, we proceed by controlling two distinct types of deformations, keeping them small. One deformation is that in the sense of Riemannian geometry between Minkowski space and (anti-)de Sitter space (originating from the cosmological term), whereas the other is that in the sense of Cartan geometry between these Klein geometries and a general Cartan geometry (originating from matter energy-momentum)\footnote{Since Cartan geometry subsumes Riemannian geometry, note that a deformation in the sense of Cartan geometry implicitly includes deformation which differs depending on whether its origin is the cosmological term or matter energy-momentum (when viewed through the lens of Cartan geometry). We treat them separately.}. Here, we shall proceed under the assumption that both of these deformations are extremely small.

To avoid unnecessary confusion, in what follows we first discuss the doformation of Klein geometry in the sense of Riemannian geomety, and subsequently address the deformation of Cartan geometry on the sense of Cartan geometry. Furthermore, for simplicity, we set $U=M$.

\subsubsection{Small deformations from Minkowski space time}
Let us consider cases in model geometries with a cosmological constant---namely de Sitter and anti-de Sitter spaces---where the deviation from Minkowski space is negligible. This corresponds to considering the situation where the cosmological constant $\Lambda$ is extremely close to zero. Rather than working directly with $\Lambda$ itself, here we formulate the argument using $l=\sqrt{\frac{3}{\lambda\Lambda}}$, or equivalently $\sqrt{\lambda\Lambda}$. Throughout this subsection, the Lie group (algebra) and geometric setups remain identical to those on the preceding subsection. In this context, algebraically rephrasing the smallness of the cosmological constant (or more precisely, $\sqrt{\lambda\Lambda}$ or $\frac{1}{l}$) implies that we consider a regime where the values of the Lie bracket among elements of $\quot{\mathfrak{g}}{\mathfrak{h}}=\quot{\so\rbra{d+1,1}}{\so\rbra{d,1}},\quot{\so\rbra{d,2}}{\so\rbra{d,1}}$ within the Lie algebra $\mathfrak{g}=\so\rbra{d+1,1},\so\rbra{d,2}$ are close to zero. Therefore, in this subsection, we expand physical quantities as instance, the Lie bracket $\sbra{\mathsf{P}_{a},\mathsf{P}_{b}}$ in Eq.~\eqref{eq:LB} can be regarded as vanishing even in the cases of de Sitter and anti-de Sitter spaces, just as in Minkowski space; as noted above, this succinctly reflects the near-zero nature of the cosmological constant from an algebraic viewpoint.

This perspective addresses the former of the two types of deformations appearing in spacetime: namely, the \emph{deformation of the model geometry} characterizing local spacetime features (parameterized by the cosmological constant), as opposed to the \emph{deformation away from the model geometry} describing global spacetime properties (parameterized by Newton's gravitational constant). Next, let us consider the latter deformation as well, and examine the higher-form symmetries present under such conditions.

\subsubsection{Linearized gravity and $1$-form symmetry}
We now describe gravity as a deformation from the space $\quot{G}{H}$ described by Klein geometry to a Cartan geometry. Specifically, we expand the Cartan connection representing gravity as a power series in the square root of Newton's constant $\sqrt{G}$, and restrict our analysis up to the first-order terms. The object under consideration is the pulled-back Cartan connection $A=\rbra{\omega,e}$ on spacetime, expanded in powers of $\sqrt{G}$ as
\begin{align}
A=\rbra*{\underline{\omega}+\sqrt{G}\varpi,\underline{e}+\sqrt{G}\varepsilon}+\CMcal{O}\rbra*{\sqrt{G}^{2}}.
\end{align}
Here, $\underline{A}\coloneqq\rbra{\underline{\omega},\underline{e}}$ denotes the background Cartan connection on $\quot{G}{H}$, while $\CMcal{A}\coloneqq\rbra{\varpi,\varepsilon}$ represents the first-order deviation in $\sqrt{G}$ from this background. Note that since Klein geometry is flat, the curvature of $\underline{A}$ vanishes. Taking this into account, the curvature $F$ of $A$ can be expanded in terms of $\sqrt{G}$ as
\begin{align}
F&=\sqrt{G}\rbra{d\CMcal{A}+\wedgespair{\underline{A}}{\CMcal{A}}}+\CMcal{O}\rbra*{\sqrt{G}^{2}}\\
&=\sqrt{G}\rbra{d\varpi+\wedgespair{\underline{\omega}}{\varpi}+\wedgespair{\underline{e}}{\varepsilon},d\varepsilon+\wedgespair{\underline{\omega}}{\varepsilon}+\wedgespair{\varpi}{\underline{e}}}+\CMcal{O}\rbra*{\sqrt{G}^{2}},
\end{align}
where$\sqrt{G}\rbra{a,b}$ means $\rbra{\sqrt{G}a,\sqrt{G}b}$. Writing the first-order term in $\sqrt{G}$ of the curvature $F$ as $\CMcal{F}$, the MacDowell-Mansouri action functional \eqref{eq:MMAF} takes the lowest-order form in $\sqrt{G}$ given by
\begin{align}
\label{eq:MMAFfLG}
S\sbra{\CMcal{A}}=-\frac{l^{2}}{2\lambda}\int_{M}\abra{\widehat{\CMcal{F}},\widehat{\CMcal{F}}}.
\end{align}
This constitutes the MacDowell-Mansouri action functional for linearized gravity.

It should be noted that the quantity $\wedgespair{\underline{e}}{\varepsilon}$ in $\widehat{\CMcal{F}}$ is the Lie bracket between $\quot{\mathfrak{g}}{\mathfrak{h}}$-valued $1$-forms, which generally consists of terms of second order or higher in $\frac{1}{l}$.

\vspace{\baselineskip}

In what follows, we shall analyze the higher-form symmetries of linearized gravity on spacetime, considering separately the case where Minkowski space serves as the model geometry with $\frac{1}{l}$ set strictly to zero, and the case where (anti-)de Sitter space serves as the model geometry with $\frac{1}{l}$ being slightly greater than zero.

\paragraph{Minkowski case}
When Minkowski space is chosen as the model geometry, the zeroth-order quantities in $\sqrt{G}$, namely $\underline{\omega}$ and $d\underline{e}$, vanish. In other words, the background Cartan connection $\underline{A}$ of the model geometry is a closed form. Under a shift transformation $\CMcal{A}\mapsto\CMcal{A}+\alpha$ by an $\iso\rbra{3,1}$-valued closed $1$-form $\alpha$, the square term $\abra{\widehat{\CMcal{F}},\widehat{\CMcal{F}}}$ of the first-order curvature $\CMcal{F}$ contained in the MacDowell-Mansouri action functional for linearized gravity \eqref{eq:MMAFfLG} generates extra terms given by
\begin{align}
\abra{d\widehat{\CMcal{A}},\widehat{\wedgespair{\underline{A}}{\alpha}}}+\abra{\widehat{\wedgespair{\underline{A}}{\alpha}},\widehat{\wedgespair{\underline{A}}{\alpha}}}.
\end{align}
Since the first term becomes an exact form due to the fact that $\underline{A}$ is closed, $\abra{\widehat{\CMcal{F}},\widehat{\CMcal{F}}}$ can be said to be invariant under the shift transformation $\CMcal{A}\mapsto\CMcal{A}+\alpha$, provided we neglect exact forms and terms quadratic in $\alpha$. Furthermore, when Minkowski space is taken as the model geometry, it is necessary to subtract the quadratic term in the Riemann curvature from Eq.~\eqref{eq:MMAFfLG}; however, in the present case, the first-order quantity in $\sqrt{G}$ of the Riemann curvature is $d\varpi$, which is likewise invariant under the transformation $\CMcal{A}\mapsto\CMcal{A}+\alpha$. Nevertheless, under this transformation, the torsion $d\varepsilon+\wedgespair{\varpi}{\underline{e}}$ is not invariant (as the second term survives)\footnote{Viewed directly in terms of $\CMcal{A}$, this means that in $\CMcal{F}=d\CMcal{A}+\wedgespair{\underline{A}}{\CMcal{A}}$, the term $\wedgespair{\underline{A}}{\alpha}$ remains. Because the spin connection conponent of $\underline{A}$ vanishes, even if $\alpha$ possesses a spin connection component, it does not affect $\widehat{\CMcal{F}}$ (the spin sonnection component of $\CMcal{F}$); however, its effect on the vielbein component does persist.}. If one wishes to avoid this discrepancy, it suggests that the shift transformation should not be taken as an $\iso\rbra{3,1}$-valued closed $1$-form, but must instead be restricted to one valued in $\quot{\iso\rbra{3,1}}{\so\rbra{3,1}}\cong\omathbb{R}^{3,1}$. Note that at this point, without neglecting the second-order terms in $\alpha$, the second term in the above equation vanishies due to $\widehat{\wedgespair{\underline{A}}{\alpha}}=0$. Consequently, linearized gravity with Minkowski space as its model geometry exhibits a shift symmetry with respect to the vielbein component of the Cartan connection. Nevertheless, if one strictly demands only the invariance of the action functional itself, it can be stated that a shift symmetry extended beyond the vielbein component is admitted.

\begin{tcolorbox}[enhanced,attach boxed title to top left={xshift=5mm,yshift=-3mm},boxed title style={colframe=black!30,colback=white},fonttitle=\bfseries,coltitle=black,colframe=black!30,colback=white,breakable=true,top=4mm,title={Linearized gravity with Minkowski space as model geometry}]
The Palatini (or MacDowell-Mansouri) action functional composed of the gravitational field $A=\underline{A}+\sqrt{G}\CMcal{A}+\CMcal{O}\rbra*{\sqrt{G}^{2}}$ on Minkowski space is invariant under the shift transformation $\CMcal{A}\mapsto\CMcal{A}+\alpha$ by an $\iso\rbra{3,1}$-valued closed $1$-form $\alpha$. However, the transformations that preserve zero torsion are restricted to those where $\alpha$ takes values in $\quot{\iso\rbra{3,1}}{\so\rbra{3,1}}$.
\end{tcolorbox}

\paragraph{(anti-)de Sitter case}
When (anti-)de Sitter space is chosen as the model geometry, the zeroth-order quantities in $\sqrt{G}$, namely $\underline{\omega}$ and $d\underline{e}$, do not vanish. In other words, the background Cartan connection $\underline{A}$ of the model geometry is no longer necessarily a closed form. Under a shift transformation $\CMcal{A}\mapsto\CMcal{A}+\alpha$ by an $\so\rbra{4,1}$-valued (or $\so\rbra{3,2}$-valued) closed $1$-form $\alpha$, the square term $\abra{\widehat{\CMcal{F}},\widehat{\CMcal{F}}}$ of the first-order curvature $\CMcal{F}$ contained in the MacDowell-Mansouri action functional for linearized gravity \eqref{eq:MMAFfLG} generates extra terms given by
\begin{align}
\abra{d\widehat{\CMcal{A}},\widehat{\wedgespair{\underline{A}}{\alpha}}}+\abra{\widehat{\wedgespair{\underline{A}}{\alpha}},\widehat{\wedgespair{\underline{A}}{\alpha}}}.
\end{align}
When the closed $1$-form $\alpha$ has only components in $\quot{\so\rbra{4,1}}{\so\rbra{3,1}}$ (or $\quot{\so\rbra{3,2}}{\so\rbra{3,1}}$), the spin connection component of $\wedgespair{\underline{A}}{\alpha}$ appearing in the first term is proportional to $\frac{1}{l^{2}}$. Therefore, for such an $\alpha$, provided we neglect exact forms as well as terms of second order in $\frac{1}{l}$, $\abra{\widehat{\CMcal{F}},\widehat{\CMcal{F}}}$ can be said to be invariant under the shift transformation $\CMcal{A}\mapsto\CMcal{A}+\alpha$. Consequently, the action functional for linearized gravity whose model geometry is an (anti-)de Sitter space very close to Minkowski space exhibits a shift symmetry with respect to the vielbein component of the Cartan connection. The curvature $\CMcal{F}$ itself develops a deviation of order $\frac{1}{l}$ in the torsion, which is its vielbein component.

\begin{tcolorbox}[enhanced,attach boxed title to top left={xshift=5mm,yshift=-3mm},boxed title style={colframe=black!30,colback=white},fonttitle=\bfseries,coltitle=black,colframe=black!30,colback=white,breakable=true,top=4mm,title={Linearized gravity with (anti-)de Sitter space as model geometry}]
The MacDowell-Mansouri action functional composed of the gravitational field $A=\underline{A}+\sqrt{G}\CMcal{A}+\CMcal{O}\rbra*{\sqrt{G}^{2}}$ on de Sitter space (or anti-de Sitter space) is invariant under the shift transformation $\CMcal{A}\mapsto\CMcal{A}+\alpha$ by a $\quot{\so\rbra{4,1}}{\so\rbra{3,1}}$-valued (or $\quot{\so\rbra{3,2}}{\so\rbra{3,1}}$-valued) closed $1$-form $\alpha$ when $l$ is sufficiently large (i.e., when considering terms up to first order in $\frac{1}{l}$).
\end{tcolorbox}

\vspace{\baselineskip}

Summarizing the cases where the model geometry of spacetime is given by Minkowski space or (anti-)de Sitter space, we conclude that when the deformation of spacetime away from the model geometry induced by matter energy-momentum is captured up to first order in $\sqrt{G}$, and the deformation of the model geometry itself away from Minkowski space induced by the cosmological constant is captured up to first order in $\sqrt{\lambda\Lambda}\propto\frac{1}{l}$, the gravitational field on the spacetime possesses a $1$-form shift symmetry with respect to the vielbein component. By decomposing the total deformation of spacetime (as described by Cartan geometry) into the deformation away from the model geometry and the deformation of the model geometry itself away from flat space (in the sense of Riemannian geometry), we have been able to uncover the existence of such a symmetry.

\paragraph{Higher-order effects of the cosmological constant}
In the preceding discussion, in parallel with considering terms up to the first order in the square root of Newton's gravitational constant (which characterized one type of deformation), we focused on the situation where the absolute value of the cosmological constant $\lambda\Lambda$ (defining another type of deformation), is sufficiently small---quantitatively considering only terms up to the first order in $\sqrt{\lambda\Lambda}$. Here, let us examine the effects of higher-order terms in $\sqrt{\lambda\Lambda}$.

We expand the zeroth-order term $\underline{A}$ and the first-order term $\CMcal{A}$ of the gravitational field $A$ with respect to $\sqrt{G}$ into power series of $\sqrt{\lambda\Lambda}$ as follows:
\begin{align}
\underline{A}=\underline{A}^{\rbra{0}}+\sqrt{\lambda\Lambda}\underline{A}^{\rbra{1}}+\sqrt{\lambda\Lambda}^{2}\underline{A}^{\rbra{2}}+\cdots,\qquad
\CMcal{A}=\CMcal{A}^{\rbra{0}}+\sqrt{\lambda\Lambda}\CMcal{A}^{\rbra{1}}+\sqrt{\lambda\Lambda}^{2}\CMcal{A}^{\rbra{2}}+\cdots.
\end{align}
Furthermore, the shift $\alpha$ acting on $\CMcal{A}$ is similarly expanded as
\begin{align}
\alpha=\alpha^{\rbra{0}}+\sqrt{\lambda\Lambda}\alpha^{\rbra{1}}+\sqrt{\lambda\Lambda}^{2}\alpha^{\rbra{2}}+\cdots.
\end{align}
Substituting these expressions into $\abra{d\widehat{\CMcal{A}},\widehat{\wedgespair{\underline{A}}{\alpha}}}$ yields
\begin{align}
\abra{d\widehat{\CMcal{A}},\widehat{\wedgespair{\underline{A}}{\alpha}}}
&=\abra{d\widehat{\CMcal{A}}^{\rbra{0}},\widehat{\wedgespair{\underline{A}^{\rbra{0}}}{\alpha^{\rbra{0}}}}}\\
&\quad+\sqrt{\lambda\Lambda}\rbra*{\abra{d\widehat{\CMcal{A}}^{\rbra{0}},\widehat{\wedgespair{\underline{A}^{\rbra{1}}}{\alpha^{\rbra{0}}}}}+\abra{d\widehat{\CMcal{A}}^{\rbra{1}},\widehat{\wedgespair{\underline{A}^{\rbra{0}}}{\alpha^{\rbra{0}}}}}+\abra{d\widehat{\CMcal{A}}^{\rbra{0}},\widehat{\wedgespair{\underline{A}^{\rbra{0}}}{\alpha^{\rbra{1}}}}}}\\
&\quad+\sqrt{\lambda\Lambda}^{2}\biggl\lparen\abra{d\widehat{\CMcal{A}}^{\rbra{0}},\widehat{\wedgespair{\underline{A}^{\rbra{2}}}{\alpha^{\rbra{0}}}}}+\abra{d\widehat{\CMcal{A}}^{\rbra{1}},\widehat{\wedgespair{\underline{A}^{\rbra{1}}}{\alpha^{\rbra{0}}}}}+\abra{d\widehat{\CMcal{A}}^{\rbra{2}},\widehat{\wedgespair{\underline{A}^{\rbra{0}}}{\alpha^{\rbra{0}}}}}+\\
&\hspace{2cm}+\abra{d\widehat{\CMcal{A}}^{\rbra{0}},\widehat{\wedgespair{\underline{A}^{\rbra{1}}}{\alpha^{\rbra{1}}}}}+\abra{d\widehat{\CMcal{A}}^{\rbra{1}},\widehat{\wedgespair{\underline{A}^{\rbra{0}}}{\alpha^{\rbra{1}}}}}+\abra{d\widehat{\CMcal{A}}^{\rbra{0}},\widehat{\wedgespair{\underline{A}^{\rbra{0}}}{\alpha^{\rbra{2}}}}}\biggr\rparen\\
&\quad+\cdots.
\end{align}
Since $\alpha$ possesses only the vielbein component, $\widehat{\wedgespair{-}{-}}$ is always proportional to $\frac{1}{l^{2}}$, i.e., $\sqrt{\lambda\Lambda}^{2}$. Therefore, if we consider terms up to the $n$-th order ($n\ge1$) in $\sqrt{\lambda\Lambda}$ for $\underline{A}$ and $\CMcal{A}$, it turns out that $\abra{d\widehat{\CMcal{A}},\widehat{\wedgespair{\underline{A}}{\alpha}}}$ exhibits invariance under shift transformations involving $\alpha^{\rbra{n-1}}$ and $\alpha^{\rbra{n}}$ (with all other components set to zero). The same reasoning applies to $\abra{\widehat{\wedgespair{\underline{A}}{\alpha}},\widehat{\wedgespair{\underline{A}}{\alpha}}}$. In particular, the case $n=1$ reproduces the exact result discussed above. In other words, even when higher-order effects of the cosmological constant cannot be neglected, a shift symmetry persists around these higher-order pertuebations. This implies that the transformations leaving the action functional invariant depend on the characteristic scale under consideration.

\subsubsection{Noether current associated with the symmetry}
Assuming, as befor, that spacetime adopts one of the three Klein geometries above as its model geometry, we shall now investigate the nature of the Noether current (conserved current) associated with the $1$-form shift symmetry identified earlier.

\vspace{\baselineskip}

Under the shift transformation $\CMcal{A}\mapsto\CMcal{A}+\alpha$ by a $\quot{\mathfrak{g}}{\so\rbra{3,1}}$-valued closed $1$-form $\alpha$ ($\mathfrak{g}=\iso\rbra{3,1},\so\rbra{4,1},\so\rbra{3,2}$) in the MacDowell-Mansouri action functional \eqref{eq:MMAFfLG}, the variation up to first order in $\alpha$ is given by
\begin{align}
S\sbra{\CMcal{A}+\alpha}-S\sbra{\CMcal{A}}=-\frac{l^{2}}{\lambda}\int_{M}\abra{\wedgespair{\alpha}{\underline{e}},d\widehat{\CMcal{A}}}.
\end{align}
Extracting $\alpha$ from this expression yields the continuity equation for the Noether current (conserved current). To obtain this, we expand the various Lie-algebra-valued differential forms using the basis $\cbra{\mathsf{M}_{ab}}_{a,b=0}^{3}$ of $\so\rbra{3,1}$ and the basis $\cbra{\mathsf{P}_{a}}_{a=0}^{3}$ of $\quot{\mathfrak{g}}{\so\rbra{3,1}}$. Performing this expansion, the variation is rewritten as
\begin{align}
S\sbra{\CMcal{A}+\alpha}-S\sbra{\CMcal{A}}=-\frac{l^{2}}{2\lambda}\int_{M}\alpha^{a}\wedge\underline{e}^{b}\wedge d\CMcal{A}^{cd}\abra{\underbrace{\sbra{\mathsf{P}_{a},\mathsf{P}_{b}}}_{\mathclap{=-\frac{\lambda}{l^{2}}\mathsf{M}_{ab}}},\mathsf{M}_{cd}}.
\end{align}
Here, $\abra{\mathsf{M}_{ab},\mathsf{M}_{cd}}=\tr{\rbra{\mathsf{M}_{ab}\rbra{{\star_{\text{int}}}\mathsf{M}_{cd}}}}$ is equal to $\epsilon_{abcd}$. Utilizing this relation as well, we obtain
\begin{align}
S\sbra{\CMcal{A}+\alpha}-S\sbra{\CMcal{A}}
&=\int_{M}\alpha^{a}\wedge\rbra*{\frac{1}{2}\epsilon_{abcd}\underline{e}^{b}\wedge d\CMcal{A}^{cd}}\\
&=\int_{M}\alpha^{a}\wedge\sbra*{d\rbra*{-\frac{1}{2}\epsilon_{abcd}\underline{e}^{b}\wedge\varpi^{cd}}-\frac{1}{2}\epsilon_{abcd}\tensor{\underline{\omega}}{^{b}_{e}}\wedge\underline{e}^{e}\wedge d\varpi^{cd}}.
\end{align}
In the intermediate steps, we used $\CMcal{A}^{cd}=\varpi^{cd}$ (since $\widehat{\CMcal{A}}$ is the $\so\rbra{3,1}$-component of $\CMcal{A}$, which coincides with $\varpi$). Now, examining the equations for the Riemann curvature and torsion associated with $\underline{e}$ and $\underline{\omega}$, namely $d\underline{\omega}+\underline{\omega}\wedge\underline{\omega}=\frac{1}{l^{2}}\underline{e}\wedge\underline{e}$ and $d\underline{e}+\underline{\omega}\wedge\underline{e}=0$, we observe that the lowest-order term of $\underline{e}$ in $\frac{1}{l}$ is $\frac{1}{l^{0}}$, whereas $\underline{\omega}$ is at least of order $\frac{1}{l}$. Therefore, retaining only up to the lowest order in $\frac{1}{l}$, the second term in the final expression can be dropped. Consequently, the $2$-form Noether current (conserved current) $J_{a}$ is determined as
\begin{align}
J_{a}={\star_{\eta}}\rbra*{\frac{1}{2}\epsilon_{abcd}\underline{e}^{b}\wedge\varpi^{cd}},
\end{align}
where ${\star_{\eta}}$ denotes the Hodge star operator defined with respect to the metric $\eta$\footnote{Since the operator acts on a quantity of first order in $\sqrt{G}$, we use ${\star_{\eta}}$ associated with $\eta$ rather than the Hodge star operator associated with the general metric pulled back by the vielbein.}. On-shell, this current satisfies $d{\star_{\eta}}J_{a}=0$ when $l$ is sufficiently large, and can therefore be regarded as approximately conserved in the regime of sufficiently large $l$.

The current obtained above formally coincides with the one obtained in Section~\ref{subsec:1FSoLGoMST}, where only Minkowski spacetime was considered. The discussion in Section~\ref{subsec:1FSoLGoMST} precisely corresponds to the limit $\frac{1}{l}\to0$ ($l\to\infty$), in which case the conservation law holds strictly.

\vspace{\baselineskip}

Following the discussion in Section~\ref{subsec:1FSoLGoMST}, let us rewrite the current obtained above as $J={\star_{\eta}}\wedgespair{\underline{e}}{{\star_{\text{int}}}\varpi}$. If this were $\wedgespair{\underline{e}}{\varpi}$, it could be rewritten as $-d_{\underline{\omega}}\varepsilon$ owing to the vanishing of the torsion (where $d_{\underline{\omega}}$ denote the exterior covariant derivative associated with $\underline{\omega}$). Up to first order in $\frac{1}{l}$, the current satisfies $d_{\underline{\omega}}{\star_{\eta}}J=0$. Although $d_{\underline{\omega}}$ is an exterior covariant derivative, since the curvature of $\underline{\omega}$ vanishes up to first order in $\frac{1}{l}$, it satisfies $d_{\underline{\omega}}^{2}=0$ within that approximation, thereby allowing us to define a cohomology. Identifying this cohomology eith the de Rham cohomology defined by the exterior derivative $d$, one can almost directly parallel the argument in Section~\ref{subsec:1FSoLGoMST}. Namely, there exists a $\quot{\mathfrak{g}}{\so\rbra{3,1}}$-valued $1$-form $\tilde{\varepsilon}$ satisfying $d_{\underline{\omega}}\tilde{\varepsilon}=\wedgespair{\underline{e}}{{\star_{\text{int}}}\varpi}$.

\vspace{\baselineskip}

Through the arguments presented above, it has been established that a shift symmetry for the linearized vielbein holds strictly in the case of Minkowski spacetime, and that an approximate shift symmetry can also be formulated for spacetimes with a non-vanishing cosmological constant, such as de Sitter and anti-de Sitter spacetime. This approximation is controlled by the cosmological constant as a parameter, becoming increasingly precise as the cosmological constant approaches zero. In other words, the degree to which this symmetry is broken provides an estimate of the value of the cosmological constant characterizing the spacetime as a maximally symmetric space (ignoring the presence of matter and other gauge fields).

\section{Conclusions and Discussions}
Based on the MacDowell-Mansouri formulation of gravity expressed in terms of Cartan geometry, we investigated the $1$-form symmetry present in linearized gravity, where quantities of first order in the square root of the Newton's gravitational constant and the cosmological constant within the Cartan connection (decomposed into spin connection and vielbein) are treated as linearized perturbations. In this framework, there exists a shift symmetry for the (linearized) Cartan connection that closely resembles the gauge field shift symmetry in Maxwell electromagnetism. Namely, we have demonstrated the existence of a $1$-form shift symmetry for the vielbein component of the Cartan connection in linearized gravity on Minkowski spacetime, as well as an approximate $1$-form shift symmetry for the vielbein component in linearized gravity on (anti-)de Sitter spacetimes with a very small cosmological constant. Furthermore, by employing a higher-order power series expansion in the cosmological constant, we have shown that the shift symmetry is replaced by a new, order-dependent shift symmetry in (anti-)de Sitter spacetimes, corresponding to the specific order of the expansion under consideration.

The $1$-form shift symmetry for the linearized gravity in the sense of the MacDowell-Mansouri formulaion---that is, for the linearized Cartan connection---does not originate from a shift of the entire Cartan connection; rather, this symmetry is restricted to the sector corresponding to translations---which can be regarded as an abelian Lie algebra---and their small deformations induced by the cosmological constant. This fact may provide an understanding of why the spacetime symmetry is reduced to the Lorentz symmetry in the MacDowell-Mansouri formulation.

While the present analysis was primarily restricted to the classical regime, we intend to extend these arguments to the quantum level in future work. Doing so will necessitate a deeper investigation into the quantum theory of linearized gravity, and pursuing a profound understanding of these quantum aspects remains a central outlook for future research.

\section*{Acknowledgment}
\addcontentsline{toc}{section}{Acknowledgment}
I would like to express my sincere gratitude to my advisor, Prof. Sakaguchi, and the members of our laboratory for organizing a seminar on generalized symmetry, which greatly helped me deepen my understanding and apply it to my research. I am also deeply grateful to Prof. Sakaguchi for carefully reading through a draft of this manuscript and checking its contents.

I would like to thank the Yukawa Institute for Theoretical Physics (YITP) at Kyoto University. Discussions during the YITP workshop YITP-W-26-08, ``Strings and Fields 2026,'' were very helpful in completing this work.

Discussions during the Physical Society of Japan 81st Annual Meeting (2026) were useful to complete this work.

\end{document}